\documentclass[twocolumn,showpacs,preprintnumbers,amsmath,amssymb,prb,superscriptaddress]{revtex4}

\usepackage{amsmath}
\usepackage{amssymb}
\usepackage{graphicx}
\usepackage{bm}
\usepackage{color}

\newcommand{\ket}[1]{|#1\rangle}
\newcommand{\braket}[1]{\langle #1 \rangle}

\newcommand{\Braket}[1]{\left\langle #1 \right\rangle}

\def\dd{\mathrm{d}}
\def\ee{\mathrm{e}}
\def\ii{\mathrm{i}}
\def\vnabla{\bm{\nabla}}

\def\rot{\vnabla\times}

\def\dgg#1{\{#1\}^{\dagger}}

\def\Hc{\mathrm{H.c.}}

\def\Im{\mathrm{Im}}

\def\half{\frac{1}{2}}

\def\die{\varepsilon}
\def\diez{\varepsilon_0}
\def\diebg{\varepsilon_{\text{bg}}}
\def\diebgex{\varepsilon_{\text{bg}}^{\text{ex}}}
\def\muz{\mu_0}

\def\wex{\omega}
\def\wbx{\varOmega}
\def\wres{\varOmega}
\def\wT{\omega_{\mathrm{T}}}
\def\DLT{\varDelta_{\mathrm{LT}}}
\def\Dbx{\varDelta_{\mathrm{bx}}}
\def\mz{m_0}
\def\mex{m_{\mathrm{ex}}}

\def\mbx{m_{\mathrm{bx}}}

\def\dampex{\gamma_{\mathrm{ex}}}
\def\dampbx{\gamma_{\mathrm{bx}}}

\def\ccxr{g}

\def\thick{d}
\def\area{S}
\def\win{\omega_{\mathrm{in}}}

\def\vkin{\bm{k}_{\mathrm{in}}}

\def\Iin{I_{\mathrm{in}}}
\def\ketg{\ket{\mathrm{g}}}

\def\keti{\ket{\mathrm{0}}}

\def\ketbx#1{\ket{#1}_{\mathrm{bx}}}
\def\ketxx#1{\ket{#1}_{\mathrm{2ex}}}

\def\expbx{\mathcal{B}}
\def\expbxw{\tilde{\mathcal{B}}}

\def\dw{\Delta\omega}

\def\intone{C^{(1)}}
\def\inttwo{C^{(2)}}
\def\intsignal{C^{(2)\text{S}}}
\def\intnoise{C^{(2)\text{N}}}

\def\css{S}

\def\mcss{\overleftrightarrow{\bm{\mathsf{S}}}}

\def\crd{W}

\def\mcrd{\overleftrightarrow{\bm{\mathsf{W}}}}

\def\wvfbx{F}

\def\rmfbx{\varPhi}

\def\cmfbx{g^{\text{bx}}}

\def\cmfex{g^{\text{ex}}}

\def\MT{$\mathrm{M_T}$}
\def\ML{$\mathrm{M_L}$}
\def\eV{\ensuremath{\mathrm{eV}}}
\def\meV{\ensuremath{\mathrm{meV}}}
\def\microeV{\ensuremath{\mu\mathrm{eV}}}

\def\microm{\ensuremath{\mu\mathrm{m}}}
\def\nm{\ensuremath{\mathrm{nm}}}

\def\ps{\ensuremath{\mathrm{ps}}}

\def\vzero{\bm{0}}
\def\vunit{\bm{e}}

\def\vunitz{\bm{e}_z}
\def\vr{\bm{r}}
\def\vrp{\bm{r}_{\parallel}}

\def\vk{\bm{k}}
\def\kp{\bar{k}}
\def\kpin{\bar{k}_\text{in}}
\def\kpone{\bar{k}_1}
\def\kptwo{\bar{k}_2}

\def\kbgex{k_{\text{bg}}^{\text{ex}}}

\def\dimP{M}
\def\vdimP{\bm{\mathcal{P}}}

\def\vdimE{\bm{\mathcal{E}}}

\def\gR{\tilde{R}}

\def\gT{\tilde{T}}

\def\gM{\tilde{M}}
\def\dG{\mathcal{G}}

\def\mdG{\overleftrightarrow{\bm{\mathcal{G}}}}

\def\DDz{\mathrm{D}_z}

\def\munit{\overleftrightarrow{\mathbf{1}}}
\def\mzero{\overleftrightarrow{\mathbf{0}}}

\def\mG{\overleftrightarrow{\bm{\mathsf{G}}}}
\def\mLN{\overleftrightarrow{\bm{\mathsf{H}}}^{\text{LN}}}
\def\mNN{\overleftrightarrow{\bm{\mathsf{H}}}^{\text{NN}}}

\def\oH{\hat{H}}
\def\oHem{\hat{H}_{\mathrm{em}}}
\def\oHint{\hat{H}_{\mathrm{int}}}

\def\oHex{\hat{H}_{\mathrm{ex}}}

\def\oHres{\hat{H}_{\mathrm{res}}}

\def\oex{\hat{b}}
\def\oexd{\hat{b}^{\dagger}}

\def\obx{\hat{B}}
\def\obxd{\hat{B}^{\dagger}}
\def\ores{\hat{d}}
\def\oresd{\hat{d}^{\dagger}}

\def\ovE{\hat{\bm{E}}}

\def\ovPex{\hat{\bm{P}}_{\mathrm{ex}}}

\def\hex{\check{b}}

\def\hexone{\check{b}^{(1)}}

\def\hbx{\check{B}}

\def\hvPex{\check{\bm{P}}_{\mathrm{ex}}}

\def\hvJz{\check{\bm{J}}_0}
\def\hE{\check{E}}
\def\hvE{\check{\bm{E}}}
\def\hvEz{\check{\bm{E}}_0}
\def\hvEin{\check{\bm{E}}_{\text{0,IN}}}
\def\hvEout{\check{\bm{E}}_{\text{0,OUT}}}

\def\hvESCT{\check{\bm{E}}_{\text{RHPS}}}

\def\hvELIN{\check{\bm{E}}_{\text{LIN}}}
\def\hvENL{\check{\bm{E}}_{\text{NL}}}
\def\hrsrc{\check{\mathcal{D}}}

\def\ovEz{\hat{\bm{E}}_0}

\begin{document}


\title{Theoretical framework of entangled-photon generation from biexcitons in nano-to-bulk crossover regime with planar geometry}

\author{Motoaki Bamba}
\altaffiliation[Present address: ]{Laboratoire Mat\'eriaux et Ph\'enom\`enes Quantiques, Universit\'e Paris Diderot-Paris 7 et CNRS, 
Case 7021, B\^atiment Condorcet, 75013 Paris, France.
E-mail: motoaki.bamba@univ-paris-diderot.fr}
\affiliation{Department of Materials Engineering Science, 
Osaka University, Toyonaka, Osaka 560-8531, Japan}
\author{Hajime Ishihara}
\affiliation{Department of Physics and Electronics, 
Osaka Prefecture University, Sakai, Osaka 599-8531, Japan}

\date{\today}

\begin{abstract}
We have constructed a theoretical framework of the biexciton-resonant
hyperparametric scattering for the pursuit of high-power and high-quality
generation of entangled photon pairs.
Our framework is applicable to nano-to-bulk crossover regime
where the center-of-mass motion of excitons and biexcitons is confined.
Material surroundings and the polarization correlation of generated
photons can be considered.
We have analyzed the entangled-photon generation from CuCl film,
by which ultraviolet entangled-photon pairs are generated,
and from dielectric microcavity embedding a CuCl layer.
We have revealed that
in the nano-to-bulk crossover regime we generally get
a high performance from the viewpoint of statistical accuracy,
and the generation efficiency can be enhanced by the optical cavity
with maintaining the high performance.
The nano-to-bulk crossover regime has a variety of degrees of freedom
to tune the entangled-photon generation,
and the scattering spectra
explicitly reflect quantized exciton-photon coupled modes
in the finite structure.
\end{abstract}

\maketitle
\section{Introduction\label{sec:intro}}
Entangled photon pairs have been discussed in relation with the Einstein-Padolsky-Rosen (EPR) paradox,\cite{einstein35}
and nowadays they play an important role in quantum information technologies.
The pursuit of their high-quality and high-efficiency generation is a fascinating subject in the fields of quantum optics and solid-state physics.
In addition to the standard generation method by
parametric down-conversion (PDC) in second-order nonlinear crystals \cite{kwiat95,kwiat99}
the generation scheme using a semiconductor quantum dot
\cite{akopian06,stevenson06,Young2009PRL,Salter2010N,Dousse2010N}
attracts much attention,
because purely a single pair of entangled photons
is created in principle,
and it can be a deterministic source of entangled pairs.
Recently, the generation efficiency is highly enhanced by implementing
an optical cavity structure with distributed Bragg reflectors (DBRs)
\cite{Young2009PRL}
and by a molecule of micropillars.\cite{Dousse2010N}
Further,
the emission by electric injection has been reported.
\cite{Salter2010N}
On the other hand, the development of entangled photons as an excitation light source is of growing importance for the next-generation technologies
of fabrication and chemical reaction.\cite{Kawabe2007OE}
For this purpose,
high-power and high-quality entangled-photon beams are absolutely necessary,
and this high-power but probabilistic generation
is another direction of research
in addition to the deterministic generation by single quantum dots.

\begin{figure}[tbp] 
\includegraphics[width=.48\textwidth]{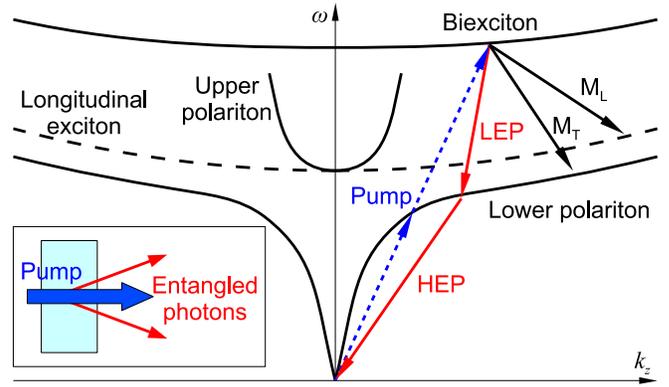}
\caption{The biexciton-resonant hyperparametric scattering (RHPS)
is depicted on dispersion curves of biexciton, exciton-polariton,
and longitudinal exciton.
Biexcitons are resonantly created by two-photon absorption,
and entangled-photon pairs are emitted
when the biexcitons decay into the lower exciton-polariton branch.
This emission appears in scattering spectrum as two peaks called LEP and HEP
(lower and higher energy polaritons) as seen in Fig.~\ref{fig:3}.
Due to the conservation of energy and wavevector,
the positions of the two peaks depend on scattering angle,\cite{ueta86bx,edamatsu04}
and the entangled two photons are emitted symmetrical about the pump beam
as shown in the inset.
Two additional peaks called $\mathrm{M_T}$ and $\mathrm{M_L}$ in scattering spectra originate from the biexciton decay
into transverse and longitudinal exciton levels, respectively.}
\label{fig:1}
\end{figure}
In the process of PDC,\cite{kwiat95,kwiat99}
an incident photon with frequency $\win$ and wavenumber $\vkin$
splits into two photons $(\omega_1,\vk_1)$ and $(\omega_2,\vk_2)$
with satisfying the conservation of energy $\win=\omega_1+\omega_2$
and of wavevector $\vkin=\vk_1+\vk_2$.
This second-order nonlinear process creates
polarization-correlated entangled-photon pairs
in nonlinear optical crystals with a birefringence.
On the other hand, Savasta et al.\cite{savasta99ssc} suggested
and Edamatsu et al.\cite{edamatsu04} experimentally demonstrated
that ultraviolet entangled-photon pairs are generated by the biexciton-resonant hyperparametric scattering (RHPS) in CuCl
(see Fig.~\ref{fig:1}).
The RHPS is a third-order nonlinear process,
in which two incident photons resonantly creates a biexciton
(excitonic molecule) with $(2\win,2\vkin)$
and it spontaneously collapses into a photon pair satisfying
$2\win=\omega_1+\omega_2$ and $2\vkin=\vk_1+\vk_2$.
Since the lowest level of biexcitons in CuCl,
which was resonantly excited in the experiment,
has zero angular momentum,\cite{ueta86bx}
the emitted pair consists of left- and right-circularly polarized photons
conserving the total angular momentum.
Owing to the two possible decay paths involving exciton-polaritons,
the emitted photons are polarization-entangled.

The generation efficiency of RHPS is much higher than that of PDC, 
because of the giant oscillator strength of the two-photon absorption
involving the biexciton.\cite{ueta86bx}
However, in the first experiment,\cite{edamatsu04}
a part of observed pairs has no entanglement,
and this noise was subtracted in the estimation of entanglement 
of the generated pairs.
As indicated by Oohata et al.,\cite{oohata07}
the main contribution of the unentangled pairs is
an accidental collapse of two biexcitons,
and this problem has been successfully suppressed
by using high-repetition and weak-power laser pulses,
because the number of unentangled pairs (noise) is increased by ${\Iin}^4$
for increasing the pumping power ${\Iin}$
while the number of entangled pairs (signal) is proportional to ${\Iin}^2$.
However, this fundamental trade-off problem between signal intensity
and $S/N$ ratio should be resolved from the improvement of material structures
\cite{Bamba2010EntangleLetter}
in addition to the improvement of pumping condition
of Ref.~\onlinecite{oohata07}.
While one solution is using a single quantum dot as a deterministic source,
\cite{akopian06,stevenson06,Young2009PRL,Salter2010N,Dousse2010N}
for the pursuit of high-power generation
there is a proposal of using an optical cavity embedding
an excitonic quantum well
for the improvement of generation efficiency.\cite{ajiki07,oka08}
Furthermore, owing to the rapid radiative decay by the exciton superradiance
(enhancement of interaction volume between excitons and photons),
\cite{ichimiya09prl,bamba09crossover}
we have theoretically revealed
that the trade-off problem can be resolved
simultaneously realizing a high generation efficiency
by using an optical cavity embedding an excitonic layer 
in nano-to-bulk crossover regime.
\cite{Bamba2010EntangleLetter}

In a microcrystal, such as quantum dot and quantum well,
smaller than the Bohr radius of excitons,
the electron and hole are individually confined in the crystal,
and the relative motion of excitons and also the binding energy
are strongly modified from those in bulk crystal.
When the crystal size is larger than the exciton Bohr radius but small enough
compared to the light wavelength, the center-of-mass motion of excitons
are confined, and the center-of-mass kinetic energy is quantized.
\cite{tredicucci93,tang95}
When the crystal size is comparable or a few times larger than the wavelength
(nano-to-bulk crossover regime),
the system is characterized by exciton-photon coupled modes with
peculiar resonance energy and radiative life time,
and the coupled modes are gradually reduced to bulk polaritons
with increasing the crystal size.
\cite{knoester92,bjork95,agranovich97,ajiki01,bamba09radlett,bamba09crossover}
In this crossover regime, the system shows
a variety of optical responses
compared to bulk materials and also to quantum dots
due to the center-of-mass confinement of excitons
and the spatially resonant coupling with electromagnetic fields.
Actually, owing to the recent development of nano-scale fabrication,
anomalous nonlinear optical processes have been reported
in semiconductor nano-structures and in the nano-to-bulk crossover regime.
\cite{ishihara96,akiyama99,ishihara01,ishihara02jul,cho03,ishihara03,ishihara04,syouji04,ishihara07,kojima08,Yasuda2009PRB,ichimiya09prl}
Further, it shows a rapid radiative decay rate of excitons
on the order of 100\;fs due to the exciton superradiance.\cite{ichimiya09prl}
Concerning the entangled photon generation,
while the performance of PDC method is almost governed
by the choice of nonlinear materials and its thickness,
the RHPS method significantly depends on 
the quantum states of excitons and biexcitons,
because it is a resonant process involving the elementary excitations.
In the nano-to-bulk crossover regime,
the generation of entangled photon pairs by RHPS
can be significantly modified with respect to
frequencies, angles, polarizations,
and phase difference of the generated entangled state
as discussed in our previous letter.\cite{Bamba2010EntangleLetter}
In the present long paper, we will show the detailed theoretical framework
for the investigation of the entangled-photon generation
in nano-to-bulk crossover regime with multilayer structures, 
especially an excitonic layer embedded in DBRs.

We explain our theoretical framework in Sec.~\ref{sec:theory},
and show in detail the way to calculate the one-photon scattering intensity
and the two-photon coincidence intensity of RHPS
in the case of multilayer structure in Sec.~\ref{ch:model}.
The calculation results are shown in Sec.~\ref{sec:results},
and the discussion is summarized in Sec.~\ref{sec:summary}.

\section{Theoretical framework}\label{sec:theory}
The emission spectra from Bose-Einstein condensation of biexcitons 
were calculated by Inoue and Hanamura,\cite{inoue76}
and they also showed the relation between energies and scattering angles
of two peaks called LEP and HEP (lower and higher energy polaritons).
Later, Hanamura and Takagahara\cite{hanamura79}
calculated line shapes of the so-called {\MT} and {\ML} peaks, 
which are emitted by the relaxations of biexcitons
to transverse and longitudinal excitons, respectively.
The entanglement of the scattered photons by RHPS was first pointed out
by Savasta et al.,\cite{savasta99ssc}
and their theoretical framework\cite{savasta99prb}
is based on the quantum electrodynamics (QED)
theory for dispersive and absorbing media \cite{huttner92,knoll01}
and on the exciton-exciton correlation functions calculated from first principles.
\cite{ostreich95,ostreich98}

In the present paper, in order to correctly treat the center-of-mass confinement
of excitons, we start from the QED theory of excitons,\cite{bamba08qed}
which simultaneously solves the equation of motion of excitons and 
of electromagnetic fields with inheriting the concepts of the above QED theories
\cite{huttner92,knoll01}
and of the semiclassical nonlocal theory\cite{cho91,cho03}
(or the so-called ABC-free theory\cite{cho86}).
It is well known that the center-of-mass motion of excitons raises more than one
propagating modes of exciton-polaritons in their band gap frequency,
and the RHPS process has been used to observe the dispersion of polaritons
\cite{inoue76,itoh77,itoh78,ueta79,ueta86bx}
and also to measure the translational masses of excitons and biexcitons.
\cite{mita80,mita80ssc,nozue82,ueta86bx}
Moreover, optical responses explicitly reflects
the confinement of center-of-mass motion of excitons
in nano-structured materials
and also in the nano-to-bulk crossover regime,
\cite{tredicucci93,tang95,ishihara96,akiyama99,ishihara01,ishihara02jul,cho03,ishihara03,ishihara04,syouji04,ishihara07,kojima08,Yasuda2009PRB,bamba09crossover,ichimiya09prl}
which we investigate in the present paper.

Concerning the treatment of biexcitons,
we suppose the excitons as pure bosons
and consider an exciton-exciton interaction leading to the creation of biexcitons.
However, instead of the detailed treatment in the theory of Savasta et al.,\cite{savasta99prb}
we simply assume the relative motion of the lowest level of biexcitons
with some parameters measured in experiments,\cite{ueta86bx,akiyama90,tokunaga99}
and the coefficients of the exciton-exciton interaction is replaced
by the wavefunction and the binding energy of biexcitons.
This treatment is very simple and useful to catch the behavior of biexciton
lowest level in CuCl even in the nano-to-bulk crossover regime,
because the exciton and biexciton states in CuCl has been
well analyzed by the bipolariton theory\cite{ivanov93,ivanov98}
and RHPS experiments.\cite{tokunaga99,tokunaga00,tokunaga01}
While the treatment of biexcitons is in general a four-body problem 
with two electrons and two holes and it is usually a hard work,
owing to the above mentioned simple treatment,
we can easily discuss the polarization correlation of photon pairs
emitted from the biexciton lowest level, which has no angular momentum.

Moreover, by the use of the dyadic Green's function for the wave equation
of electric field, we can consider the surroundings of excitonic material,
such as an optical cavity consisting of two DBRs.
In order to extract the scattering fields,
instead of using the input-output relation,
\cite{matloob95,savasta96,gruner96aug,knoll01,savasta02josab,khanbekyan03}
we consider the definition of Green's function 
and commutation relations of fluctuation operators.
This simple treatment is valid at least in multilayer systems
and useful to consider complicated structures.

In the following subsections, we show our theoretical framework
to calculate the signal and noise intensities by RHPS.
We show the Hamiltonian in Sec.~\ref{sec:hamiltonian},
and the equations of motion are derived in Sec.~\ref{sec:eq_motion}.
In order to discuss the RHPS, we use some approximations, which are explained
in Sec.~\ref{sec:approx_RHPS}.
The model of biexcitons are shown in Sec.~\ref{sec:wvfbx}.
In order to solve the equations of motion, we use the Green's function
technique explained in Sec.~\ref{sec:Green}.
Finally, we derive the expression of observables in Sec.~\ref{sec:nonlinear}.

\subsection{Hamiltonian} \label{sec:hamiltonian}
Our theoretical framework is based on the QED theory of excitons.\cite{bamba08qed}
The Hamiltonian is written as
\begin{equation}
\oH = \oHex + \oHres + \oHint + \oHem,
\end{equation}
where $\oHex$ describes the excitonic system,
$\oHres$ represents a reservoir for the nonradiative damping of excitons,
$\oHint$ is the exciton-photon interaction,
and $\oHem$ describes the electromagnetic fields and a background dielectric
medium as discussed in Ref.~\onlinecite{suttorp04} 
and also used in Ref.~\onlinecite{bamba08qed}.
In order to discuss the biexciton-associated RHPS,
we consider an exciton-exciton interaction with coefficient $V_{\mu,\nu;\mu',\nu'}$. Namely, the Hamiltonian of excitonic system is written as
\begin{equation} \label{eq:def-Hamilt-ex} 
\oHex = \sum_{\mu} \hbar\wex_{\mu} \oexd_{\mu} \oex_{\mu}
+ \half \sum_{\mu,\mu',\nu,\nu'} V_{\mu,\nu;\mu',\nu'}
  \oexd_{\mu} \oexd_{\nu} \oex_{\nu'} \oex_{\mu'},
\end{equation}
where $\oex_{\mu}$ is the annihilation operator of an exciton
in state $\mu$ and $\wex_{\mu}$ is its eigenfrequency.
We treat the excitons as pure bosons satisfying
\begin{subequations}
\begin{align}
\left[ \oex_{\mu}, \oexd_{\mu'} \right] & = \delta_{\mu,\mu'}, \\
\left[ \oex_{\mu}, \oex_{\mu'} \right] & = 0,
\end{align}
\end{subequations}
and their non-bosonic behavior is described 
by the exciton-exciton interaction, the second term
in Eq.~\eqref{eq:def-Hamilt-ex}.
The reservoir $\oHres$ is written as
\begin{align}
\oHres & = \sum_{\mu} \int_0^{\infty}\dd\wres\ \bigl\{
    \hbar\wres \oresd_{\mu}(\wres) \ores_{\mu}(\wres)
\nonumber \\ & \quad
+ \left[ \oex_{\mu} + \oexd_{\mu} \right]
  \left[ \ccxr_{\mu}(\wres) \ores_{\mu}(\wres)
       + \ccxr^*_{\mu}(\wres) \oresd_{\mu}(\wres) \right] \bigr\},
\end{align}
where $\ores_{\mu}(\wres)$ is the annihilation operator of harmonic
oscillator with frequency $\wres$
interacting with excitons in state $\mu$,
and $\ccxr_{\mu}(\wres)$ is the coupling coefficient.
The oscillators are independent with each other
and satisfy the following commutation relations:
\begin{subequations} \label{eq:[ores,ores]} 
\begin{align}
[ \ores_{\mu}(\wres), \oresd_{\mu'}(\wres') ]
& = \delta_{\mu,\mu'} \delta(\wres-\wres'), \\
[ \ores_{\mu}(\wres), \ores_{\mu'}(\wres') ]
& = 0.
\end{align}
\end{subequations}
Further, $\oHint$ is simply written as
a product of electric field $\ovE(\vr)$ and excitonic polarization
$\ovPex(\vr)$:
\begin{equation}
\oHint = - \int\dd\vr\ \ovPex(\vr) \cdot \ovE(\vr).
\end{equation}
Here, the excitonic polarization is represented as
\begin{equation}
\ovPex(\vr) = \sum_{\mu} \vdimP_{\mu}(\vr) \oex_{\mu} + \Hc,
\end{equation}
where the coefficient $\vdimP_{\mu}(\vr)$ is expressed by the exciton
center-of-mass wavefunction $\cmfex_{\mu}(\vr)$ and 
unit vector $\vunit_{\mu}$ of polarization direction as
\begin{equation} \label{eq:vdimP} 
\vdimP_{\mu}(\vr) = \dimP \vunit_{\mu} \cmfex_{\mu}(\vr).
\end{equation}
The absolute value of $\dimP$ can be evaluated 
by the longitudinal-transverse (LT) splitting energy
$\DLT = |\dimP|^2/\diez\diebgex$ of excitons, the vacuum permittivity $\die_0$,
and the background dielectric constant $\diebgex$ of the excitonic medium.

\subsection{Equations of motion} \label{sec:eq_motion}
According to Ref.~\onlinecite{bamba08qed}
or the QED theories of dispersive and absorbing media,
\cite{huttner92,knoll01,suttorp04}
the equation of motion of electric field $\ovE(\vr)$ is derived 
in frequency domain as
\begin{align}& \label{eq:Maxwell-E-Jz-Pex} 
\rot\rot\hvE^+(\vr,\omega)
- \frac{\omega^2}{c^2}\diebg(\vr,\omega)\hvE^+(\vr,\omega)
\nonumber \\ &
= \ii\muz\omega\hvJz(\vr,\omega) + \muz\omega^2\hvPex^+(\vr,\omega).
\end{align}
Here, $\mu_0$ is the vacuum permeability,
$\diebg(\vr,\omega)$ is the dielectric function
of the background medium with arbitrary three dimensional structure.
We write an operator with a check ( $\check{}$ ) in frequency domain.
$\hvJz(\vr,\omega)$ describes the fluctuation
of electromagnetic fields and satisfies
\begin{align} \label{eq:[hvJz,hvJz]} 
&   \left[ \hvJz(\vr,\omega), \dgg{\hvJz(\vr',{\omega'}^*)} \right]
= \left[ \hvJz(\vr,\omega), \hvJz(\vr',-\omega') \right]
\nonumber \\ &
= \delta(\omega-\omega') \delta(\vr-\vr') \frac{\diez\hbar\omega^2}{\pi}
    \Im[\diebg(\vr,\omega)] \munit.
\end{align}
In the same manner as in Ref.~\onlinecite{bamba08qed},
we obtain the equation of excitons' motion in frequency domain as
\begin{align}& \label{eq:motion-hex} 
\left[ \hbar\wex_{\mu}-\hbar\omega-\ii\dampex/2 \right]
\hex_{\mu}(\omega)
\nonumber \\ &
= \int\dd\vr\ \vdimP_{\mu}^*(\vr) \cdot \hvE^+(\vr,\omega)
  + \hrsrc_{\mu}(\omega)
\nonumber \\ & \quad
- \sum_{\nu} \sum_{\mu',\nu'} V_{\mu,\nu;\mu',\nu'}
  \int_{-\infty}^{\infty}\dd t\ \frac{\ee^{\ii\omega t}}{2\pi}
  \oexd_{\nu}(t) \oex_{\nu'}(t) \oex_{\mu'}(t),
\end{align}
where $\dampex$ is the nonradiative damping width
defined in terms of $\{\ccxr_{\mu}(\wres)\}$ as shown in Eq.~(D7) 
of Ref.~\onlinecite{bamba08qed},
and $\hrsrc_{\mu}(\omega)$ represents the fluctuation by the damping satisfying
\begin{align}& \label{eq:[hrsrc,hrsrc]} 
\left[ \hrsrc_{\mu}(\omega), \dgg{\hrsrc_{\mu'}({\omega'}^*)} \right]
= \left[ \hrsrc_{\mu}(\omega), \hrsrc_{\mu'}(-\omega') \right]
\nonumber \\ &
= \delta_{\mu,\mu'} \delta(\omega-\omega') \frac{\hbar\dampex}{2\pi}.
\end{align}
The last term on the right hand side of Eq.~\eqref{eq:motion-hex}
is the nonlinear term due to the exciton-exciton interaction.

Here, we define a new operator
\begin{equation} \label{eq:def-obx} 
\obx_{\lambda}
\equiv \half \sum_{\mu,\nu} \wvfbx_{\lambda,\mu,\nu}^* \oex_{\nu} \oex_{\mu},
\end{equation}
which annihilates a biexciton (excitonic molecule) in state $\lambda$
and describes a two-exciton eigen state $\obxd_{\lambda}\ketg$
by applying it to the ground state $\ketg$ of matter system.
The coefficient $\wvfbx_{\lambda,\mu,\nu}$ is invariant 
by the exchange of two exciton indices as
\begin{equation} \label{eq:invariant-mu-nu} 
\wvfbx_{\lambda,\mu,\nu} = \wvfbx_{\lambda,\nu,\mu}.
\end{equation}
Further, it is ortho-normal
\begin{equation}
\half\sum_{\mu,\nu} \wvfbx_{\lambda,\mu,\nu} \wvfbx_{\lambda',\mu,\nu}^*
= \delta_{\lambda,\lambda'},
\end{equation}
and also has a completeness
\begin{equation} \label{eq:complete-wvfbx} 
\sum_{\lambda} \wvfbx_{\lambda,\mu,\nu}
\wvfbx_{\lambda,\mu',\nu'}^*
= \delta_{\mu,\mu'} \delta_{\nu,\nu'}
+ \delta_{\mu,\nu'} \delta_{\nu,\mu'}.
\end{equation}
From the excitonic Hamiltonian $\oHex$ [Eq.~\eqref{eq:def-Hamilt-ex}],
the coefficient $\wvfbx_{\lambda,\mu,\nu}$ 
and eigen frequency $\wbx_{\lambda}$
of biexciton eigen state $\lambda$
should satisfy
\begin{equation} \label{eq:wvfbx-satisfy-1} 
( \hbar\omega_{\mu} + \hbar\omega_{\nu} ) \wvfbx_{\lambda,\mu,\nu}
+ \sum_{\mu',\nu'} V_{\mu,\nu;\mu',\nu'} \wvfbx_{\lambda,\mu',\nu'}
= \hbar\wbx_{\lambda} \wvfbx_{\lambda,\mu,\nu}.
\end{equation}
By using Eqs.~\eqref{eq:invariant-mu-nu} and \eqref{eq:complete-wvfbx},
we can rewrite Eq.~\eqref{eq:def-obx} as
\begin{equation}
\sum_{\lambda} \wvfbx_{\lambda,\mu,\nu} \obx_{\lambda}
= \oex_{\nu} \oex_{\mu}.
\end{equation}
Therefore, from this relation and Eq.~\eqref{eq:wvfbx-satisfy-1},
we can rewrite Eq.~\eqref{eq:motion-hex} as
\begin{widetext}
\begin{align} \label{eq:motion-hex-2} 
\left[ \hbar\wex_{\mu}-\hbar\omega-\ii\dampex/2 \right]
\hex_{\mu}(\omega)
& = \int\dd\vr\ \vdimP_{\mu}^*(\vr) \cdot \hvE^+(\vr,\omega)
  + \hrsrc_{\mu}(\omega)
\nonumber \\ & \quad
+ \sum_{\lambda,\nu}( \hbar\wex_{\mu}+ \hbar\wex_{\nu}
                     - \hbar\wbx_{\lambda})
  \wvfbx_{\lambda,\mu,\nu}
  \int_{-\infty}^{\infty}\dd\omega'\
  \dgg{\hex_{\nu}(\omega'-\omega)} \hbx_{\lambda}(\omega').
\end{align}
\end{widetext}
On the other hand, by deriving the equation of motion for $\obx_{\lambda}$
and by using the above relations, we get
\begin{align}& \label{eq:motion-hbx} 
(\hbar\wbx_{\lambda}-\hbar\omega) \hbx_{\lambda}(\omega)
\nonumber \\ &
= \sum_{\mu,\nu}\wvfbx_{\lambda,\mu,\nu}^*\ \int_{-\infty}^{\infty}\dd\omega'\
    (\hbar\wex_{\nu}-\hbar\omega') \hex_{\nu}(\omega')
    \hex_{\mu}(\omega-\omega').
\end{align}
In principle, the biexciton RHPS process is described
by the three equations of motion \eqref{eq:Maxwell-E-Jz-Pex}, \eqref{eq:motion-hex-2}, 
and \eqref{eq:motion-hbx},
and commutation relations \eqref{eq:[hvJz,hvJz]} and \eqref{eq:[hrsrc,hrsrc]}.
However, in the actual calculation, we use the following approximation.

\subsection{Approximation for RHPS process} \label{sec:approx_RHPS}
We suppose that a coherent light beam resonantly excites biexcitons
and their amplitude is large enough compared to the vacuum fluctuation.
In this situation, if we do not consider the other higher processes,
the biexciton operator in the nonlinear term of Eq.~\eqref{eq:motion-hex-2}
can be replaced by its amplitude
$\mathcal{B}_{\lambda}(\omega') = \braket{\hbx_{\lambda}(\omega')}$.
Further, we replace $\hex_{\nu}(\omega'-\omega)$
in the nonlinear term by $\hexone_{\nu}(\omega'-\omega)$,
which satisfies the linear equation
\begin{align}& \label{eq:motion-hexone} 
\left[ \hbar\wex_{\mu}-\hbar\omega-\ii\dampex/2 \right]
\hexone_{\mu}(\omega)
\nonumber \\ &
= \int\dd\vr\ \vdimP_{\mu}^*(\vr) \cdot \hvE^+(\vr,\omega)
  + \hrsrc_{\mu}(\omega).
\end{align}
Simultaneously solving this equation and Eq.~\eqref{eq:Maxwell-E-Jz-Pex},
$\hexone_{\mu}(\omega)$ can be expressed by
the fluctuation operators $\hvJz(\vr,\omega)$ and $\hrsrc_{\mu}(\omega)$.
Its calculation is straightforward
by using the Green's function technique
as will be shown in section \ref{sec:Green}.
Under the above approximation, Eq.~\eqref{eq:motion-hex-2} is rewritten as
\begin{widetext}
\begin{align} \label{eq:motion-hex-3} 
\left[ \hbar\wex_{\mu}-\hbar\omega-\ii\dampex/2 \right]
\hex_{\mu}(\omega)
&
\simeq \int\dd\vr\ \vdimP_{\mu}^*(\vr) \cdot \hvE^+(\vr,\omega)
  + \hrsrc_{\mu}(\omega)
\nonumber \\ & \quad
+ \sum_{\lambda,\nu}( \hbar\wex_{\mu}+ \hbar\wex_{\nu}
                     - \hbar\wbx_{\lambda})
  \wvfbx_{\lambda,\mu,\nu}
  \int_{-\infty}^{\infty}\dd\omega'\
  \dgg{\hexone_{\nu}(\omega'-\omega)} \mathcal{B}_{\lambda}(\omega').
\end{align}
\end{widetext}
By solving this equation and Eq.~\eqref{eq:Maxwell-E-Jz-Pex},
we can represent $\hvE^+(\vr,\omega)$
by the fluctuation operators $\hvJz(\vr,\omega)$ and $\hrsrc_{\mu}(\omega)$.
This calculation is also straightforward by using the Green's function.

For the calculation of $\mathcal{B}_{\lambda}(\omega)$,
we suppose that the biexciton amplitude is not decreased by the scattering,
because its contribution is small compared to the pumping light.
Under this approximation,
by phenomenologically introducing a damping constant $\dampbx$,
the biexciton amplitude is obtained from Eq.~\eqref{eq:motion-hbx} as
\begin{align}& \label{eq:<hbx>=<hexone>2} 
\mathcal{B}_{\lambda}(\omega)
\simeq \frac{1}{\hbar\wbx_{\lambda}-\hbar\omega-\ii\dampbx/2}
  \sum_{\mu,\nu}\wvfbx_{\lambda,\mu,\nu}^*\
\nonumber \\ & \quad \times
  \int_{-\infty}^{\infty}\dd\omega'\
    (\hbar\wex_{\nu}-\hbar\omega') \braket{\hexone_{\nu}(\omega')}
    \braket{\hexone_{\mu}(\omega-\omega')},
\end{align}
where $\braket{\hexone_{\nu}(\omega')}$ can be calculated
from Eqs.~\eqref{eq:Maxwell-E-Jz-Pex} and \eqref{eq:motion-hexone}
by considering an incident light beam
as a homogeneous solution of Eq.~\eqref{eq:Maxwell-E-Jz-Pex}.
Under the weak bipolariton regime,\cite{oka08}
where the coupling between exciton-polariton and biexciton
is small enough compared to their broadening,
the approximated expression \eqref{eq:<hbx>=<hexone>2}
of biexciton amplitude is sufficient for the discussion of RHPS process.
While Savasta et al.~considered the equation of motion of projection operators,
\cite{savasta99prb,savasta99ssc}
they also used a similar approximation for the treatment of biexcitons
under the detailed verification of its validity.

\subsection{Model of biexcitons} \label{sec:wvfbx}
Although $\wvfbx_{\lambda,\mu,\nu}$ and $\wbx_{\lambda}$ should be in principle
determined from Eq.~\eqref{eq:wvfbx-satisfy-1} for given nonlinear coefficient 
$V_{\mu,\nu;\mu',\nu'}$, we instead express $\wvfbx_{\lambda,\mu,\nu}$ and 
$\wbx_{\lambda}$ by using experimental results.
This treatment is useful because we already know many parameters
of the lowest level of biexcitons in CuCl
by the longstanding experimental and theoretical studies.\cite{ueta86bx}

It is well known that the lowest level of biexcitons in CuCl
is singlet and has zero angular momentum,
because of the exchange interactions 
between two electrons and between two holes.\cite{ueta86bx}
Since we suppose the resonant two-photon excitation of the lowest level,
we only consider the lowest relative motion of biexciton in our calculation.
Further, according to the RHPS experiments in Ref.~\onlinecite{tokunaga99},
the lowest biexciton state mainly consists of $1s$ excitons,
and the contribution from the higher exciton states was estimated
in the order of $10^{-4}$.
Therefore, we consider only $1s$ relative motion of excitons,
which has a degree of freedom of polarization direction $\xi_{\mu}=\{x,y,z\}$.
According to the exciton and biexciton states in bulk CuCl,\cite{ueta86bx}
the lowest biexciton level $\ketbx{J=0,M=0}$ with zero angular momentum
is represented  as
\begin{align} \label{eq:|J=0>=|xx>} 
\ketbx{J=0,M=0}
& = \frac{1}{2} \{ \ketxx{0,0;0,0} + \ketxx{1,1;1,-1}
\nonumber \\ & \quad \quad
                    + \ketxx{1,-1;1,1} - \ketxx{1,0;1,0} \},
\end{align}
where $\ketxx{j_1,m_1;j_2,m_2}$ is the two-exciton state
represented in terms of angular momenta $(j_1,m_1)$ and $(j_2,m_2)$
of two excitons.
This expression surely reflects the polarization correlation of
photon pairs observed in RHPS experiments\cite{edamatsu04,oohata07}
and also determines the phase between the two states
\begin{subequations} \label{eq:Phi+} 
\begin{align}
\Phi_+ &= \left( \ket{L,R} + \ket{R,L} \right)/\sqrt{2} \\
&= \left( \ket{H,H} + \ket{V,V} \right)/\sqrt{2}.
\label{eq:Phi+=HH+VV} 
\end{align}
\end{subequations}
Here, $\ket{L,R}$ means that one photon is left- and the other is 
right-circularly polarized, and $\ket{R,L}$ is the opposite state.
$\ket{H,H}$ and $\ket{V,V}$ respectively means 
that both photons are horizontally and vertically polarized.
By rewriting each exciton state
in terms of the polarization direction as
\begin{equation}
\ket{j,m} = \begin{cases}
\ket{1,1} = - \left( \ket{x} + \ii \ket{y} \right) / \sqrt{2}, \\
\ket{1,0} = \ket{z}, \\
\ket{1,-1} = \left( \ket{x} - \ii \ket{y} \right) / \sqrt{2},
\end{cases}
\end{equation}
Eq.~\eqref{eq:|J=0>=|xx>} is rewritten as
\begin{align}& \label{eq:|J=0>=|xyz>} 
\ketbx{J=0,M=0}
\nonumber \\ & \quad
= \frac{1}{2} \{ \ketxx{0,0;0,0} - \ketxx{x;x}
- \ketxx{y;y} - \ketxx{z;z} \},
\end{align}
which also reflects the polarization correlation
\eqref{eq:Phi+=HH+VV}.

Considering the relative motion $\varPsi(\vr)$ of two excitons
in the lowest biexciton level, the coefficient is written as
\begin{equation}
\wvfbx_{\lambda,\mu,\nu} = \delta_{\lambda,\mu,\nu}
\int\dd\vr\int\dd\vr'\ \varPsi(\vr') \cmfbx_{\lambda}(\vr)
  {\cmfex_{\mu}}^*(\vr+\vr') {\cmfex_{\nu}}^*(\vr),
\end{equation}
where $\cmfex_m(\vr)$ and $\cmfbx_l(\vr)$ are 
center-of-mass wavefunctions of excitons and biexcitons,
respectively, and
\begin{equation}
\delta_{\lambda,\mu,\nu} = \delta_{\xi_{\mu},\xi_{\nu}}
\end{equation}
represents the polarization selection rule reflecting
the lowest state of biexciton [Eq.~\eqref{eq:|J=0>=|xyz>}].
Here, we suppose that the Bohr radius of biexciton
(1.5\;nm in CuCl)\cite{Singh1998PSS}
is much smaller than the crystal size,
and the relative motion of biexcitons is not strongly modified
from the bulk one.
Namely, we approximate the above expression as
\begin{equation} \label{eq:mwvfbx} 
\wvfbx_{\lambda,\mu,\nu} \simeq \delta_{\lambda,\mu,\nu}
\rmfbx \int\dd\vr\ \cmfbx_{\lambda}(\vr)\
  {\cmfex_{\mu}}^*(\vr)\ {\cmfex_{\nu}}^*(\vr),
\end{equation}
where $\rmfbx$ is defined as
\begin{equation} \label{eq:rmfbx} 
\rmfbx \equiv \int\dd\vr\ \varPsi(\vr).
\end{equation}
$|\rmfbx|^2$ represents the effective volume of the lowest biexciton state.
It was estimated by an experiment,\cite{akiyama90}
and was also used as a parameter in a theoretical work.\cite{matsuura95}

\subsection{Green's function technique} \label{sec:Green}
Here, we explain how we simultaneously solve the equation of motion 
of electric field [Eq.~\eqref{eq:Maxwell-E-Jz-Pex}]
and that of excitons [Eq.~\eqref{eq:motion-hexone}
or Eq.~\eqref{eq:motion-hex-3}].
By using the dyadic Green's function satisfying
\begin{equation} \label{eq:satisfied-mG} 
\rot\rot\mG(\vr,\vr',\omega)
- \frac{\omega^2}{c^2}\diebg(\vr,\omega)\mG(\vr,\vr',\omega)
= \delta(\vr-\vr')\munit,
\end{equation}
we can rewrite Eq.~\eqref{eq:Maxwell-E-Jz-Pex} as
\begin{equation} \label{eq:E=Ez+G*Pex} 
\hvE^+(\vr,\omega) = \hvEz^+(\vr,\omega)
  + \muz\omega^2 \int\dd\vr'\ \mG(\vr,\vr',\omega) \cdot \hvPex^+(\vr',\omega),
\end{equation}
where $\hvEz^+(\vr,\omega)$ represents the electric field 
in the background ($\oHem$) system, and it is defined as
\begin{equation} \label{eq:def-hvEz-hvJz} 
\hvEz^{+}(\vr,\omega) \equiv \ii\muz\omega
  \int\dd\vr'\ \mG(\vr,\vr',\omega) \cdot \hvJz(\vr',\omega).
\end{equation}
From Eq.~\eqref{eq:[hvJz,hvJz]}, $\hvEz^+(\vr,\omega)$ satisfies
\cite{bamba08qed}
\begin{align}& \label{eq:[Ez(r,w),Ez(r,w)]} 
  \left[ \hvEz^+(\vr,\omega), \hvEz^-(\vr',\omega') \right]
= \left[ \hvEz^+(\vr,\omega), \hvEz^+(\vr',-\omega') \right] \nonumber \\ &
= \delta(\omega-\omega') \frac{\muz\hbar\omega^2}{\ii2\pi}
  [\mG(\vr,\vr',\omega)-\mG^*(\vr,\vr',\omega)].
\end{align}
The expression of $\mG(\vr,\vr',\omega)$ in planar system
(dielectric multilayer) is already known\cite{chew95}
and will be shown in Sec.~\ref{ch:model}.

Substituting Eq.~\eqref{eq:E=Ez+G*Pex} into Eq.~\eqref{eq:motion-hex-3},
we obtain the simultaneous equation set for exciton operators 
under the rotating-wave approximation (RWA) as
\begin{widetext}
\begin{align} \label{eq:self-consist-nl} 
\sum_{\mu'} \css_{\mu,\mu'}(\omega) \hex_{\mu'}(\omega)
& = \int\dd\vr\ \vdimP_{\mu}^*(\vr) \cdot \hvEz^+(\vr,\omega)
+ \hrsrc_{\mu}(\omega)
\nonumber \\ & \quad
+ \sum_{\lambda,\nu}
  (\hbar\wex_{\mu}+\hbar\wex_{\nu}-\hbar\varOmega_{\lambda})
  \wvfbx_{\lambda,\mu,\nu}
  \int_{-\infty}^{\infty}\dd\omega'\
  \dgg{\hexone_{\nu}(\omega'-\omega)} \mathcal{B}_{\lambda}(\omega'),
\end{align}
where the coefficient on the left-hand side is defined as
\begin{align} \label{eq:def-css} 
&\css_{\mu,\mu'}(\omega)
\equiv \left[ \hbar\wex_{\mu} - \hbar\omega
            - \ii\dampex/2 \right] \delta_{\mu,\mu'}
- \muz\omega^2\int\dd\vr\int\dd\vr'\ \vdimP_{\mu}^*(\vr)
    \cdot \mG(\vr,\vr',\omega) \cdot \vdimP_{\mu'}(\vr').
\end{align}
\end{widetext}
The last term of Eq.~\eqref{eq:def-css} represents the self energy
due to the retarded interaction through the electromagnetic fields
and the longitudinal Coulomb interaction.
Further, Eq.~\eqref{eq:motion-hexone} for $\hexone_{\mu}(\omega)$
in the linear regime is also rewritten as
\begin{equation} \label{eq:self-consistent-linear} 
\sum_{\mu'} \css_{\mu,\mu'}(\omega) \hexone_{\mu'}(\omega)
= \int\dd\vr\ \vdimP_{\mu}^*(\vr) \cdot \hvEz^+(\vr,\omega)
+ \hrsrc_{\mu}(\omega).
\end{equation}
This simultaneous linear equation set is solved
by calculating the inverse matrix $\mcrd(\omega) = [\mcss(\omega)]^{-1}$,
and the commutation relation of $\hexone_{\mu}(\omega)$ is derived
in Ref.~\onlinecite{bamba08qed} as
\begin{subequations} \label{eq:[hexone,hexone]} 
\begin{align}&
\left[ \hexone_{\mu}(\omega), \dgg{\hexone_{\mu'}({\omega'}^*)} \right]
\nonumber \\ & \quad
= \delta(\omega-\omega') \frac{\hbar}{\ii2\pi}
    \left[ \crd_{\mu,\mu'}(\omega) - \crd_{\mu',\mu}^*(\omega) \right], \\
& \left[ \hexone_{\mu}(\omega), \hexone_{\mu'}(-\omega') \right]
= 0.
\end{align}
\end{subequations}
Further, Eq.~\eqref{eq:self-consist-nl} is rewritten as
\begin{align} \label{eq:self-consist-app} 
\hex_{\mu}(\omega)
&\simeq \hexone_{\mu}(\omega)
+ \sum_{\mu',\lambda,\nu} \crd_{\mu,\mu'}(\omega)
  (\hbar\wex_{\mu'}+\hbar\wex_{\nu}-\hbar\varOmega_{\lambda})
\nonumber \\ & \quad \times
  \wvfbx_{\lambda,\mu',\nu}
  \int_{-\infty}^{\infty}\dd\omega'\
  \dgg{\hexone_{\nu}(\omega'-\omega)} \mathcal{B}_{\lambda}(\omega'),
\end{align}
and, by substituting this into Eq.~\eqref{eq:E=Ez+G*Pex},
the electric field involved with RHPS is expressed under the RWA
and the approximations discussed in Sec.~\ref{sec:approx_RHPS} as
\begin{equation} \label{eq:E=Ez+E*b+ENL} 
\hvE^+(\vr,\omega) \simeq \hvEz^+(\vr,\omega)
  +  \sum_{\mu} \vdimE_{\mu}(\vr,\omega) \hexone_{\mu}(\omega)
  + \hvENL(\vr,\omega),
\end{equation}
where
\begin{equation}
\vdimE_{\mu}(\vr,\omega) 
\equiv \muz\omega^2 \int\dd\vr'\ \mG(\vr,\vr',\omega) \cdot \vdimP_{\mu}(\vr'),
\end{equation}
\begin{align} \label{eq:hvE=hvEone+hvENL} 
\hvENL^+(\vr,\omega)
& = \sum_{\mu,\mu',\lambda,\nu} \vdimE_{\mu}(\vr,\omega) \crd_{\mu,\mu'}(\omega)
  (\hbar\wex_{\mu'}+\hbar\wex_{\nu}-\hbar\varOmega_{\lambda})
\nonumber \\ & \quad \times
  \wvfbx_{\lambda,\mu',\nu}
  \int_{-\infty}^{\infty}\dd\omega'\
  \dgg{\hexone_{\nu}(\omega'-\omega)} \mathcal{B}_{\lambda}(\omega').
\end{align}

\subsection{Input and output fields} \label{sec:nonlinear}
Here, we must pay attention to the electric field $\hvEz^+(\vr,\omega)$
in the background system, which represents not only the field
from matter to an observing point $\vr$
but also the field from $\vr$ to the matter.
This means that the latter contribution must be removed from
Eq.~\eqref{eq:E=Ez+E*b+ENL} to calculate observables,
while the other terms involving $\hexone_{\mu}(\omega)$ and $\hvENL(\vr,\omega)$
represents the components emitted from the matter.
While such a calculation has been usually treated by the input-output relations,
\cite{knoll01,matloob95,savasta96,gruner96aug,savasta02josab,khanbekyan03}
we use the following treatment based on the dyadic Green's function
$\mG(\vr,\vr',\omega)$ for $\hvEz^+(\vr,\omega)$.

We separate $\hvEz^+(\vr,\omega)$
into an input field $\hvEin^+(\vr,\omega)$ from $\vr$ to the matter
and an output field $\hvEout^+(\vr,\omega)$ from the matter to $\vr$ as
\begin{equation} \label{eq:hvEz=hvEin+hvEout} 
\hvEz^+(\vr,\omega) = \hvEin^+(\vr,\omega) + \hvEout^+(\vr,\omega).
\end{equation}
By considering the causality,
the output field at time $t$ should have a correlation 
only with fields at $t' < t$,
and the commutation relation should be written as
\begin{align}
&   \left[ \hvEout^+(\vr,\omega), \hvEz^-(\vr',\omega') \right] \nonumber \\
& = \frac{1}{(2\pi)^2} \int_{-\infty}^{\infty}\dd t'\ \int_{t'}^{\infty}\dd t\
    \ee^{\ii\omega t - \ii\omega't'}
    \left[ \ovEz(\vr,t), \ovEz(\vr',t') \right] \nonumber \\
& = \delta(\omega-\omega') \frac{\muz\hbar\omega^2}{\ii2\pi}
    \mG(\vr,\vr',\omega),
\label{eq:[hvEout+,hvEz-]} 
\end{align}
where we use the fact that the dyadic Green's function
$\mG(\vr,\vr',\omega)$ satisfying Eq.~\eqref{eq:satisfied-mG}
is the retarded correlation function of the electric field:
\cite{abrikosov75ch6,bamba08qed}
\begin{align}& \label{eq:G=<[Ez,Ez]>} 
- \muz\omega^2 \mG(\vr,\vr',\omega) \nonumber \\ &
= \frac{1}{\ii\hbar} \int_{t'}^{\infty}\dd t\ \ee^{\ii\omega(t-t')}
  \Braket{\left[ \ovEz(\vr,t), \ovEz(\vr',t') \right]}.
\end{align}
In the same manner, 
the input field at $t$ should have a correlation only with fields
at $t' > t$, and the commutation relation is derived as
\begin{align}
&   \left[ \hvEin^+(\vr,\omega), \hvEz^-(\vr',\omega') \right] \nonumber \\
& = \frac{1}{(2\pi)^2} \int_{-\infty}^{\infty}\dd t'\ \int_{-\infty}^{t'}\dd t\
    \ee^{\ii\omega t - \ii\omega't'}
    \left[ \ovEz(\vr,t), \ovEz(\vr',t') \right] \nonumber \\
& = - \delta(\omega-\omega') \frac{\muz\hbar\omega^2}{\ii2\pi}
    \mG^*(\vr,\vr',\omega).
\label{eq:[hvEin+,hvEz-]} 
\end{align}
Actually, Eqs.~\eqref{eq:[hvEout+,hvEz-]} and \eqref{eq:[hvEin+,hvEz-]}
reproduces Eq.~\eqref{eq:[Ez(r,w),Ez(r,w)]}.
By using the output field $\hvEout^+(\vr,\omega)$ in the background system,
we define the scattering field excluding the input one as
\begin{align} \label{eq:def-hvESCT} 
\hvESCT^+(\vr,\omega)
& \equiv \hvE^+(\vr,\omega) - \hvEin^+(\vr,\omega) \nonumber \\
& = \hvELIN^+(\vr,\omega) + \hvENL^+(\vr,\omega),
\end{align}
where $\hvELIN^+(\vr,\omega)$ is the linear component of the electric field
excluding the input field as
\begin{equation} \label{eq:def-hvELIN} 
\hvELIN^+(\vr,\omega)
= \hvEout^+(\vr,\omega)
+ \sum_{\mu} \vdimE_{\mu}(\vr,\omega) \hexone_{\mu}(\omega).
\end{equation}
By deriving commutation relations of $\hvELIN(\vr,\omega)$ and
$\hvENL(\vr,\omega)$ from Eqs.~\eqref{eq:[Ez(r,w),Ez(r,w)]},
\eqref{eq:[hexone,hexone]}, and \eqref{eq:[hvEout+,hvEz-]},
we can evaluate the observables of RHPS.

\section{Practical calculation}\label{ch:model}
\begin{figure}[tbp] 
\includegraphics[width=.48\textwidth]{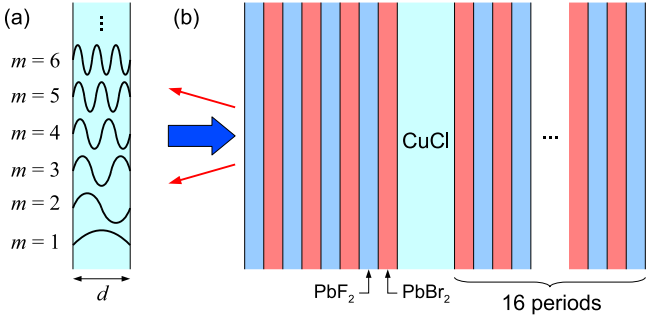}
\caption{(a) Center-of-mass wavefunctions of excitons and biexcitons
in a CuCl film. Simple sinusoidal functions vanishing at surfaces are supposed.
(b) Cavity structure considered in Figs.~\ref{fig:9} and \ref{fig:10}.
The Bragg mirrors consist of $\mathrm{PbBr_2}$ and $\mathrm{PbF_2}$.
On the transmission side, a high reflectance is achieved
by a mirror with 16 periods to suppress the leakage of photons
in this direction.
On the incident side, only 4 periods are supposed to guarantee
rapid radiative decay of entangled photons.}
\label{fig:2}
\end{figure}
Next, we apply the theoretical framework discussed in the previous section
into multilayer systems embedding a CuCl layer,
and derive expressions of one-photon scattering
intensity and two-photon coincidence intensity by RHPS.
An incident light beam propagates along $z$ axis
(perpendicular to the surface),
and photon pairs emitted into $x-z$ plane are considered
(in-plane vector is in $x$ direction).
We suppose that center-of-mass motions of excitons and biexcitons are confined
in the CuCl layer with thickness $\thick$ existing at $0 < z < \thick$.
Since we consider a large enough thickness $\thick$
compared to the Bohr radii of exciton (0.7\;nm) and biexciton (1.5\;nm),
\cite{Singh1998PSS}
the relative motions of excitons and biexcitons are not strongly modified
from the ones in bulk crystals,
and all the information of the relative motions are described
by factors $M$ and $\rmfbx$
in Eqs.~\eqref{eq:vdimP} and \eqref{eq:mwvfbx}.
As seen in Fig.~\ref{fig:2}(a),
the center-of-mass wavefunctions of excitons and biexcitons
are expanded by a series of
sinusoidal functions as
\begin{equation} \label{eq:cmf} 
\cmfbx_{\kp,m}(\vr) = \cmfex_{\kp,m}(\vr) 
= \theta(z) \frac{\ee^{\ii\kp x}}{\sqrt{\area}}\sqrt{\frac{2}{\thick}} \sin(q_mz),
\end{equation}
where $\theta(z)$ is unity for $0 < z < \thick$ and zero otherwise,
$\kp$ is the in-plane wavenumber,
$\area$ is the normalization area in $x-y$ plane,
and $q_m = m\pi/\thick$ is the confinement wavenumber in $z$ direction
for $m=1,\ 2,\ \ldots$.
We consider $\diebg(\vr,\omega)$ as a discontinuous step-like function
in $z$ direction representing the background dielectric constant
in each layer, and it does not depend on $\omega$ nor $\vrp$.
In the case of multilayer structure in Fig.~\ref{fig:2}(b),
$\diebg(\vr,\omega)$ gives the background dielectric constant $\diebgex$
for excitons at CuCl layer,
otherwise it gives the dielectric constant of each layer.
According to Ref.~\onlinecite{chew95},
if $z'$ is in the CuCl layer,
the dyadic Green's function satisfying Eq.~\eqref{eq:satisfied-mG} is expressed as
\begin{align} \label{eq:mG-3D} 
\mG_{\kp}(z,z',\omega)
& \equiv \int\dd\vrp \int \dd\vrp'\
  \frac{\ee^{-\ii\kp(x-x')}}{\area} \mG(\vr,\vr',\omega)
\nonumber \\
& = - \frac{1}{\ii2\kbgex} 
    \left[ \mdG^V_{\kp}(z,z',\omega) + \mdG^H_{\kp}(z,z',\omega) \right]
\nonumber \\ & \quad
- \frac{\vunitz\vunitz}{\diebgex\omega^2/c^2} \delta(z-z'),
\end{align}
where ${\kbgex}^2 = \diebgex\omega^2/c^2-{\kp}^2$
and $\vunit_{\xi}$ is the unit vector in $\xi$ direction.
When $z$ is in layer $j$ with dielectric constant $\die_i$,
the tensors in Eq.~\eqref{eq:mG-3D} are written as
\begin{equation} \label{eq:mdGs} 
\mdG^V_{\kp}(z,z',\omega) = \vunit_y\vunit_y \dG^V_{\kp}(z,z',\omega),
\end{equation}
\begin{equation} \label{eq:mdGp} 
\mdG^H_{\kp}(z,z',\omega) = 
\begin{pmatrix}
\DDz \DDz' & 0 & \ii \DDz \kp \\
0 & 0 & 0 \\
- \ii \kp\DDz' & 0 & {\kp}^2
\end{pmatrix} \frac{\dG^H_{\kp}(z,z',\omega)}{\sqrt{\diebgex\die_j}\omega^2/c^2},
\end{equation}
where $\DDz \equiv \partial/\partial z$ and
$\DDz' \equiv \partial/\partial z'$.
Eqs.~\eqref{eq:mdGs} and \eqref{eq:mdGp} respectively
describe the propagation of V- and H-polarized fields,
and, according to Ref.~\onlinecite{chew95},
$\dG^{V/H}_{\kp}(z,z')$ is expressed as follows.
\begin{subequations} \label{eq:dG} 
When $0 < z < d$ is in the CuCl layer,
\begin{align}
&\dG^{V/H}_{\kp}(z,z')
\nonumber \\ &
= \ee^{\ii\kbgex|z-z'|}
+ \ee^{\ii\kbgex z} \gR^{V/H}_L \left[ \ee^{\ii\kbgex z'}
  + \gR^{V/H}_R \ee^{\ii\kbgex(2\thick-z')} \right] \gM^{V/H}
\nonumber \\ & \quad
+ \ee^{-\ii\kbgex(z-\thick)} \gR^{V/H}_R \left[ \ee^{\ii\kbgex(\thick-z')}
  + \gR^{V/H}_L \ee^{\ii\kbgex(\thick+z')} \right] \gM^{V/H}.
\end{align}
When $z$ is in the leftmost semi-infinite region,
\begin{equation}
\dG^{V/H}_{\kp}(z,z')
= \ee^{- \ii k_L z} \gT^{V/H}_L \left[
    \ee^{\ii \kbgex z'}
  + \gR^{V/H}_R \ee^{\ii\kbgex(2\thick-z')}
  \right] \gM^{V/H}.
\end{equation}
When $z$ is in the rightmost semi-infinite region,
\begin{equation}
\dG^{V/H}_{\kp}(z,z')
= \ee^{\ii k_R z} \gT^{V/H}_R \left[
    \ee^{\ii\kbgex(\thick-z')}
  + \gR^{V/H}_L \ee^{\ii\kbgex(\thick+z')}
  \right] \gM^{V/H}.
\end{equation}
\end{subequations}
Here, $\gR^{V/H}_{L(R)}$ represents the generalized reflection coefficient
for $V/H$-polarized field from the CuCl layer against the left(right)-hand
neighboring,
and $\gT^{V/H}_{L(R)}$ is the generalized transmission coefficient
from the CuCl layer to the left(right)most region.
The derivation of these coefficients is shown in Ref.~\onlinecite{chew95}.
Further, ${k_{L(R)}}^2 = \die_{L(R)}\omega^2/c^2-{\kp}^2$
is the wavenumber in the left(right)most region with dielectric constant 
$\die_{L(R)}$, and the factor is defined as
$\gM^{V/H} = [ 1 - \gR^{V/H}_L \gR^{V/H}_R \ee^{\ii2\kbgex\thick} ]^{-1}$.

From Eqs.~\eqref{eq:cmf} and \eqref{eq:mG-3D}, 
we can evaluate the coefficient matrix $\mcss(\omega)$
[Eq.~\eqref{eq:def-css}]
and numerically calculate the inverse matrix
$\mcrd(\omega) = [\mcss(\omega)]^{-1}$.
From Eq.~\eqref{eq:self-consistent-linear}, the amplitude of excitons
is obtained in linear regime by
\begin{equation}
\braket{\hexone_{\mu}(\omega)}
= \sum_{\mu'} \crd_{\mu,\mu'}(\omega)
  \int\dd\vr\ \vdimP_{\mu'}^*(\vr) \cdot \braket{\hvEz^+(\vr,\omega)}.
\end{equation}
Here, $\braket{\hvEz^+(\vr,\omega)}$ represents the amplitude of electric field
in the background dielectric system $\oHem$, and can be derived
by the standard transfer matrix method\cite{chew95}
in the case of dielectric multilayers.
For simplicity, we consider a monochromatic laser light with frequency $\win$
with in-plane wavenumber $\kpin$.
Concerning the pump power $\Iin$
($\braket{\hvEz^+} \propto \sqrt{\Iin}$),
there is a scaling law for the intensity
of entangled photons as explained below.
In the present paper, since we only consider the 1$s$ exciton
and the lowest biexciton level,
the exciton states are labeled by polarization direction $\xi = \{x, y, z\}$,
in-plane wavenumber $\kp$, and index of center-of-mass motion $m$
as $\mu = \{\xi, \kp, m\}$,
and the biexciton states are labeled by $\lambda = \{\kp, m\}$.
Considering the conservation of energy and in-plane wavevector,
the amplitude of biexcitons is evaluated by Eq.~\eqref{eq:<hbx>=<hexone>2}
and we write it as
\begin{equation}
\expbx_{\kp,m}(\omega) =
\delta_{\kp,2\kpin}\delta(\omega-2\win)\expbxw_{2\kpin,m}(2\win).
\end{equation}
Further, the linear and nonlinear components
of the scattering field [Eqs.~\eqref{eq:def-hvELIN} and \eqref{eq:hvE=hvEone+hvENL}]
are simply rewritten as
\begin{widetext}
\begin{equation}
\hvE_{\text{LIN},\kp}^+(z,\omega)
= \frac{1}{\sqrt{\area}} \int\dd\vrp\ \ee^{-\ii \kp x} \hvELIN^+(\vr,\omega)
= \hvE_{0,\text{OUT},\kp}^+(z,\omega)
+ \sum_{\xi,m} \vdimE_{\xi,\kp,m}(z,\omega) \hexone_{\xi,\kp,m}(\omega),
\end{equation}
\begin{align}
\hvE_{\text{NL},\kp}^+(z,\omega)
= \frac{1}{\sqrt{\area}} \int\dd\vrp\ \ee^{-\ii \kp x} \hvENL^+(\vr,\omega)
= \sum_{\xi,m}\vdimE_{\xi,\kpin,\kp,m}^{\text{NL}}(z,\win,\omega)
  \dgg{\hexone_{\xi,2\kpin-k,m}(2\win-\omega)},
\end{align}
where the coefficients are evaluated by the following quantities and functions
\begin{align}
\vdimE_{\xi,\kp,m}(z,\omega) 
\equiv \muz\omega^2 M \sqrt{\frac{1}{\thick}} \int\dd z'\ \mG_{\kp}(z,z',\omega) \cdot \vunit_{\xi}
\sin(q_mz') \theta(z'),
\end{align}
\begin{align}
& \vdimE_{\xi,\kpin,\kp,m}^{\text{NL}}(z,\win,\omega)
= \sum_{\xi',\xi'',m',m'',n} \vdimE_{\xi'',\kp,m''}(z,\omega)
  \crd_{\{\xi'',\kp,m''\},\{\xi',\kp,m'\}}(\omega)
\nonumber \\ & \quad \times
  (\hbar\wex_{\xi',\kp,m'}+\hbar\wex_{\xi,2\kpin-\kp,m}-\hbar\varOmega_{2\kpin,n})
  \wvfbx_{\{2\kpin,n\},\{\xi',\kp,m'\},\{\xi,2\kpin-\kp,m\}}
   \tilde{\mathcal{B}}_{2\kpin,n}(2\win),
\end{align}
\begin{align}
\wvfbx_{\{2\kpin,n\},\{\xi',\kp,m'\},\{\xi,\kp',m\}}
= \delta_{\xi,\xi'} \delta_{\kp',2\kpin-\kp} \rmfbx
  \left(\frac{2}{\thick}\right)^{3/2}
  \int\dd z\ \theta(z) \sin(q_nz) \sin(q_mz) \sin(q_{m'}z),
\end{align}
\begin{align}
\css_{\{\xi,\kp,m\},\{\xi',\kp',m'\}}(\omega)
& = \left[ \hbar\wex_{\xi,\kp,m} - \hbar\omega - \ii\dampex/2 \right]
  \delta_{\xi,\xi'} \delta_{\kp,\kp'} \delta_{m,m'}
\nonumber \\ & \quad
- \delta_{\kp,\kp'} \muz\omega^2|M|^2 \int\dd z\int\dd z'\ \theta(z) \theta(z')
  \vunit_{\xi} \cdot \mG_{\kp}(z,z',\omega) \cdot \vunit_{\xi'}
  \sin(q_mz) \sin(q_{m'}z').
\end{align}
Further, from the commutation relations \eqref{eq:[hexone,hexone]}
and \eqref{eq:[hvEout+,hvEz-]}, the following relations are derived
for $\omega > 0$ and $\omega' > 0$ as
\begin{subequations} \label{eq:[hvELIN,hvENL]} 
\begin{align}&
\left[ \hvE_{\text{LIN},\kp}^+(\vr,\omega), 
       \hvE_{\text{NL},\kp'}^+(\vr',\omega') \right]
= \delta_{\kp',2\kpin-\kp} \delta(\omega+\omega'-2\win)
  \mLN_{\kpin,\kp}(z,z',\win,\omega), \\
& \left[ \hvE_{\text{LIN},\kp}^+(\vr,\omega), 
         \hvE_{\text{NL},\kp'}^-(\vr',\omega') \right]
= \mzero,
\end{align}
\end{subequations}
\begin{subequations} \label{eq:[hvENL,hvENL]} 
\begin{align}
& \left[ \hvE_{\text{NL},\kp}^-(\vr,\omega), 
       \hvE_{\text{NL},\kp'}^+(\vr',\omega') \right]
= \delta_{\kp,\kp'} \delta(\omega-\omega') \mNN_{\kpin,\kp}(z,z',\win,\omega), \\
& \left[ \hvE_{\text{NL},\kp}^+(\vr,\omega), 
       \hvE_{\text{NL},\kp'}^+(\vr',\omega') \right]
= \mzero,
\end{align}
\end{subequations}
where the tensors are defined as
\begin{align} \label{eq:mLN} 
& \mLN_{\kpin,\kp}(z,z',\win,\omega)
\equiv \frac{\hbar}{\ii2\pi} \sum_{\xi,\xi',m,m'}
  \vdimE_{\xi,\kp,m}(z,\omega) \crd_{\{\xi,\kp,m\},\{\xi',\kp,m'\}}(\omega)
  \vdimE^{\text{NL}}_{\xi',\kpin,2\kpin-\kp,m'}(z',\win,2\win-\omega),
\end{align}
\begin{align} \label{eq:mNN} 
\mNN_{\kpin,\kp}(z,z',\win,\omega)
& \equiv \frac{\hbar}{\ii2\pi} \sum_{\xi,\xi',m,m'}
  \vdimE^{\text{NL}*}_{\xi,\kpin,\kp,m}(z,\omega)
  \left[ \crd_{\{\xi,2\kpin-\kp,m\},\{\xi',2\kpin-\kp,m'\}}(2\win-\omega)
\right. \nonumber \\ & \quad \left.
       - \crd_{\{\xi',2\kpin-\kp,m'\},\{\xi,2\kpin-\kp,m\}}^*(2\win-\omega) \right] 
  \vdimE^{\text{NL}}_{\xi',\kpin,\kp,m'}(z',\win,\omega).
\end{align}

From these commutation relations, we calculate the one-photon scattering intensity
and the two-photon coincidence intensity.
When the background field $\hvE_{0,\kp}(z)$ is in vacuum state
in the scattering direction determined by $\kp$ 
and only has the quantum fluctuation,
we obtain the following relations for the initial condition $\keti$
without considering the perturbation by the exciton-exciton scattering:
\begin{align} \label{eq:hvE*ketz=0} 
& \hvE_{0,\kp}^+(z,\omega) \keti
= \hvE_{0,\text{OUT},\kp}^+(z,\omega) \keti
= \hvE_{\text{LIN},\kp}^+(z,\omega) \keti
= \hvE_{\text{NL},\kp}^-(z,\omega) \keti = \vzero.
\end{align}
When we measure the one-photon scattering intensity
in the direction $\kp$ at position $z$ with polarization direction $\xi$
and frequency $\omega$ by resolution $\dw$, the intensity is written as
\begin{align} \label{eq:intone} 
\intone_{\xi,\kpin,\kp}(z,\win,\omega)
&
= \int_{\omega-\dw/2}^{\omega+\dw/2}\dd\omega'\dd\omega''\
\braket{ \hE_{\text{RHPS},\kp,\xi}^-(z,\omega')
         \hE_{\text{RHPS},\kp,\xi}^+(z,\omega'')}
\nonumber \\ &
= \dw \left[ \mNN_{\kpin,\kp}(z,z,\win,\omega) \right]_{\xi,\xi},
\end{align}
where $\hE^{\pm}_{\text{RHPS},\kp,\xi}$ is the $\xi$ component of
$\hvE^{\pm}_{\text{RHPS},\kp}$ and
$[ \cdots ]_{\xi,\xi'}$ extracts $(\xi,\xi')$ component of the tensor.
Here, it is worth noting that 
this one-photon scattering intensity is proportional to ${\Iin}^2$,
the square of the pump power, reflecting the power dependence
of the biexciton creation. Further,
the $z$ dependence of this function
only represents the scattering direction to left or right hand side,
if the leftmost and rightmost regions are non-absorptive.
When we measure the two-photon coincidence
between the scattering fields of $(\xi_1,\kpone,z_1,\omega_1)$
and $(\xi_2,\kptwo,z_2,\omega_2)$, the intensity is calculated by
\begin{align}&
\inttwo_{\xi_1,\xi_2,\kpin,\kpone,\kptwo}(z_1,z_2,\win,\omega_1,\omega_2)
= \int_{\omega_1-\dw/2}^{\omega_1+\dw/2}\dd\omega_1'\dd\omega_1''
  \int_{\omega_2-\dw/2}^{\omega_2+\dw/2}\dd\omega_2'\dd\omega_2''
\nonumber \\ & \quad \times
\braket{ \hE_{\text{RHPS},\kpone,\xi_1}^-(z_1,\omega_1')
         \hE_{\text{RHPS},\kptwo,\xi_2}^-(z_2,\omega_2')
         \hE_{\text{RHPS},\kptwo,\xi_2}^+(z_2,\omega_2'')
         \hE_{\text{RHPS},\kpone,\xi_1}^+(z_1,\omega_1'') }.
\end{align}
By using the above commutation relations, we finally get
\begin{align} \label{eq:inttwo} 
& \inttwo_{\xi_1,\xi_2,\kpin,\kpone,\kptwo}(z_1,z_2,\win,\omega_1,\omega_2)
\nonumber \\
& = \delta_{\kptwo,2\kpin-\kpone} \tilde{\delta}(\omega_1+\omega_2,2\win)
  \intsignal_{\xi_1,\xi_2,\kpin,\kpone}(z_1,z_2,\win,\omega_1)
+ \intnoise_{\xi_1,\xi_2,\kpin,\kpone,\kptwo}(z_1,z_2,\win,\omega_1,\omega_2)
\nonumber \\ &
+ (\dw)^2 \delta_{\kpone,\kptwo} \tilde{\delta}(\omega_1,\omega_2)
  \left[ \mNN_{\kpin,\kpone}(z_1,z_2,\omega_1) \right]_{\xi_1,\xi_2}
  \left[ \mNN_{\kpin,\kptwo}(z_2,z_1,\omega_2) \right]_{\xi_2,\xi_1}.
\end{align}
\end{widetext}
Here, the function $\tilde{\delta}(\omega,\omega')$ gives unity
for $\omega \simeq \omega'$ and zero otherwise.
The first term represents the signal intensity,
i.e., the number of correlated photon pairs,
which satisfies the energy conservation $\omega_1 + \omega_2 \simeq 2\win$
by resolution $\dw$, and the intensity is calculated as
\begin{align}
&\intsignal_{\xi_1,\xi_2,\kpin,\kpone}(z_1,z_2,\win,\omega_1)
\nonumber \\ &
\equiv (\dw)^2 \left|\left[ 
{\mLN}_{\kpin,2\kpin-\kpone}(z_2,z_1,\win,2\win-\omega_1)
\right]_{\xi_2,\xi_1}\right|^2.
\end{align}
This expression is invariant for swapping the two observing conditions,
and it is also proportional to ${\Iin}^2$,
because an entangled-photon pair is emitted from a biexciton.
On the other hand, 
the second term in Eq.~\eqref{eq:inttwo} has a finite value 
for arbitrary pair of $\omega_1$ and $\omega_2$,
and represents the accidental coincidence of emitted photons
from independent two biexcitons,
because this is just the product of two one-photon scattering intensities as
\begin{align}
& \intnoise_{\xi_1,\xi_2,\kpin,\kpone,\kptwo}(z_1,z_2,\win,\omega_1,\omega_2)
\nonumber \\ &
\equiv \intone_{\xi_1,\kpin,\kpone}(z_1,\win,\omega_1)
       \intone_{\xi_2,\kpin,\kptwo}(z_2,\win,\omega_2)
\end{align}
This is also invariant for swapping the two observing conditions,
and proportional to ${\Iin}^4$.
The third term in Eq.~\eqref{eq:inttwo} represents the interference
between the two observing point, 
and has a value only for $\omega_1 \simeq \omega_2$.
Therefore, we neglect this term in the following discussion.

According to Sec.~3.10 in Ref.~\onlinecite{ueta86bx},
we suppose the translational masses of excitons and biexcitons are,
respectively, $\mex = 2.3\mz$ and $\mbx = 2.3\mex$, 
where $\mz$ is the free electron mass.
These masses were measured by RHPS experiments.\cite{mita80,mita80ssc,ueta86bx}
However, in our calculation, 
we do not consider the difference of the mass of longitudinal excitons
from that of transverse one.
From Sec.~3.2 in Ref.~\onlinecite{ueta86bx},
the transverse exciton energy at band edge 
is $\hbar\wT = 3.2022~\eV$,
LT splitting energy is $\DLT = 5.7~\meV$,
and background dielectric constant of CuCl is $\diebgex = 5.59$.
Further, according to Sec.~3.7 in Ref.~\onlinecite{ueta86bx},
the binding energy of biexciton lowest level is $\Dbx = 32.2~\meV$.
The energy of excitons including the center-of-mass kinetic energy is written as
\begin{equation}
\hbar\wex_{\xi,\kp,m}
= \hbar\wT + \frac{\hbar^2}{2\mex}\left( {\kp}^2 + {q_{m}}^2 \right).
\end{equation}
The energy of biexciton is
\begin{equation}
\hbar\wbx_{\kp,m} = 2\hbar\wT - \Dbx
  + \frac{\hbar^2}{2\mbx}\left( {\kp}^2 + {q_{m}}^2 \right).
\end{equation}
We use the other biexciton parameters reported in Ref.~\onlinecite{akiyama90}:
The phenomenological damping width is
$\dampbx = \hbar / 50~\ps = 13.2~\microeV$,
and the effective volume is
$|\rmfbx|^2 = (4000/2) \times (0.541~\nm)^3/4 = 80~\nm^3$,
where 0.541~nm is the lattice constant of CuCl,
and 4000 is a parameter representing the nonlinear strength.
In most of all calculations, 
we consider the exciton nonradiative damping width as $\dampex = 0.5~\meV$.

Because of the translational symmetry in $x-y$ plane,
the in-plane wavenumber of the system is conserved.
In the following discussion, 
we suppose that the pump field is perpendicular to the layers,
and biexcitons have zero in-plane wavenumber.
Then, a scattered photon with $\kp$ makes a pair with the one having $-\kp$.
However, their frequencies are different in general
satisfying the energy conservation $\omega_1+\omega_2 = 2\win$.
In the present paper, we define the scattering angle $\theta$ as
$\kp = (\wT/c) \sin\theta$, 
which is approximately equal to the scattering angle in vacuum.

\section{Results} \label{sec:results}
By using the theoretical framework discussed in the previous sections,
we calculate the scattering spectra by bulk crystal and by thin film
in Sec.~\ref{sec:thoer_valid}.
In Sec.~\ref{sec:perform}, we discuss the difference of entangled photon
generation by thin film from that by bulk crystal,
and show the thickness dependence of generation efficiency and performance
by RHPS in CuCl.
Finally, we discuss the the generation from a DBR cavity embedding a CuCl layer
in Sec.~\ref{sec:PRL}.

\subsection{Scattering spectra\label{sec:thoer_valid}}
\begin{figure}[tbp] 
\includegraphics[width=.48\textwidth]{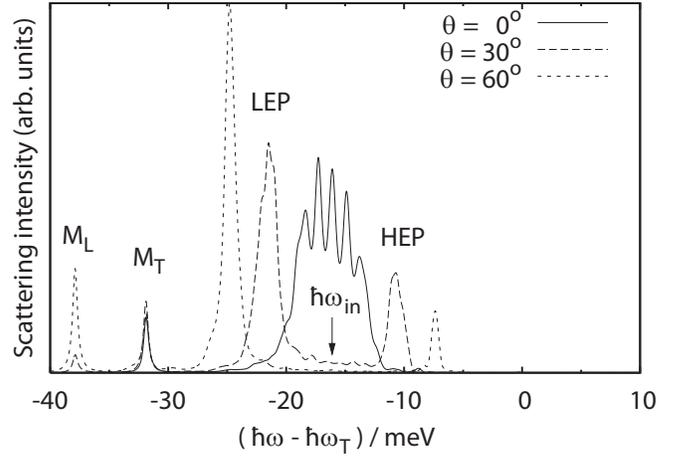}
\caption{Forward scattering spectra from a CuCl film 
with thickness $\thick=7~\microm$.
The film exists in vacuum, and the pump beam is perpendicular to it.
The pump frequency $\win$ corresponds to the two-photon absorption
due to the biexciton. The results for scattering angles $\theta=0^{\circ}$,
$30^{\circ}$, and $60^{\circ}$ are plotted with different lines
as functions of scattering frequency $\omega$.}
\label{fig:3}
\end{figure}
Fig.~\ref{fig:3} shows forward (transmission side)
scattering spectra of RHPS 
from a CuCl film with thickness $\thick=7~\microm$. We plot 
$\intone_{\xi,\kpin=0,\kp=(\wT/c)\sin\theta}(z>\thick,\win,\omega)$
as a function of $\omega$ for scattering angles
$\theta = 0^{\circ}$, $30^{\circ}$, and $60^{\circ}$,
and the spectra are summed for all the polarization direction $\xi=\{x,y,z\}$.
The CuCl film exists in vacuum,
and the pump frequency is tuned to the two-photon absorption
involving biexcitons as $\hbar\win\simeq\hbar\wT-\Dbx/2$.
Actually, $\hbar\win$ is not exactly $\hbar\wT-\Dbx/2$,
because we must also consider the phase-matching condition
(wavevector conservation)
between two polaritons and a biexciton.\cite{ueta86bx}
Since the shapes of scattering spectra do not depend on
the input power $\Iin$,
we plot the scattering intensity with arbitrary units.
The decay paths of biexcitons are depicted in Fig.~\ref{fig:1}.
As seen in Fig.~\ref{fig:3}, at $\theta = 0^{\circ}$, we have multiple peaks
at $\hbar\omega-\hbar\wT \simeq -\Dbx/2=-16.1~\meV$
and a single peak at $\hbar\omega-\hbar\wT \simeq -\Dbx = 32.2~\meV$.
The latter is called {\MT} peak, 
which is emitted by the biexciton relaxation into transverse exciton level
(exciton-like polariton).\cite{ueta86bx}
The remaining polariton with frequency $\omega\simeq\wT$ propagates backward, 
but it cannot go outside the film because of the absorption.
On the other hand, the multiple peaks at $\hbar\omega-\hbar\wT \simeq -16.1~\meV$
originate from the biexciton relaxation
into two polaritons,
and the peak structure is due to the interference inside the film
with $\thick=7\;\microm$.
Increasing the scattering angle,
the entangled peaks are split into lower and higher energy sides
satisfying the energy and wavevector conservations 
as discussed in Ref.~\onlinecite{inoue76}.
These peaks are the LEP and HEP,
and the intensity of HEP is usually smaller than that of LEP,
because of the strong absorption near the bare exciton energy $\wT$.
The angle dependence of the peak positions
obeys the discussion of Ref.~\onlinecite{inoue76}.
The peak at $\hbar\omega-\hbar\wT\simeq-\Dbx-\DLT$
is called {\ML}, which is emitted by the biexciton relaxation
into longitudinal exciton state.
The emitted photon cannot go outside when $\theta = 0^{\circ}$
because it is polarized in $z$ direction (longitudinal),
and the remaining exciton also cannot go outside due to the strong absorption
even for $\theta > 0^{\circ}$.

\begin{figure}[tbp] 
\includegraphics[width=.48\textwidth]{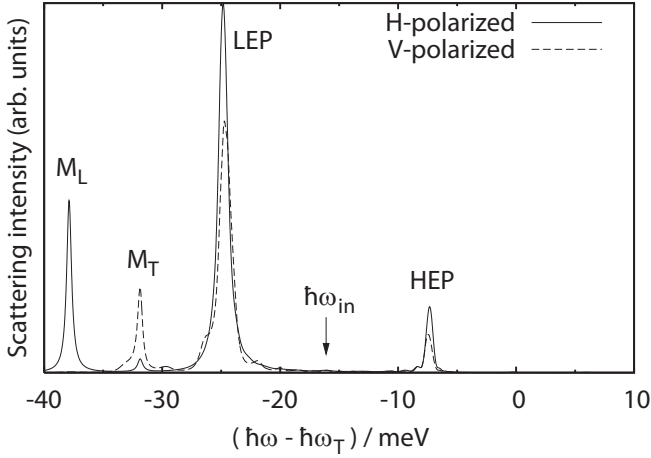}
\caption{The scattering spectra 
of horizontal(H)- and vertical(V)-polarizations are shown.
The film thickness is $\thick=7~\microm$,
and the scattering angle is $\theta=60^{\circ}$.
The other parameters are the same as in Fig.~\ref{fig:3}.}
\label{fig:4}
\end{figure}
Fig.~\ref{fig:4} shows the polarization-resolved
scattering spectra.
The film thickness is also $\thick=7\;\microm$
and the scattering angle is $\theta=60^{\circ}$.
H and V represent the horizontal and vertical polarizations, respectively,
with respect to the scattering plane.
The {\ML} peak consists of only H-polarized light,
because V-polarization does not contain the longitudinal component.
Concerning the LEP and HEP peaks,
these scattering intensities depend on the polarization.
We generally get this behavior at a non-zero scattering angle,
because the reflectance at the surface is in general different
for the two polarizations.
When we resolve the spectra with circular polarizations,
the spectra of left- and right-polarizations are the same
for any scattering angles and frequencies.

\begin{figure}[tbp] 
\includegraphics[width=.48\textwidth]{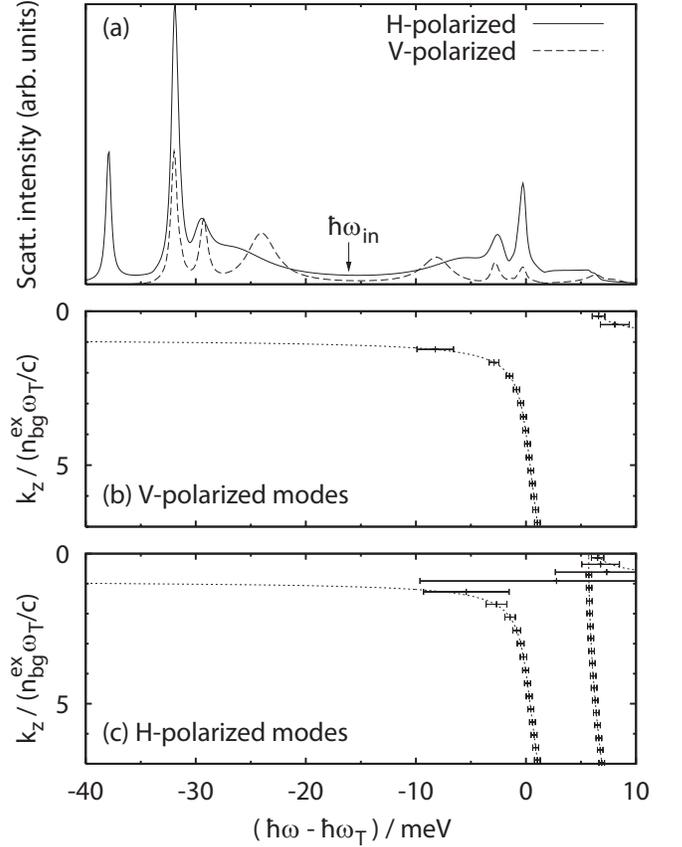}
\caption{(a) Polarization-resolved scattering spectra
for thickness $\thick=200\;\nm$ and scattering angle $\theta=60^{\circ}$.
The film exists in vacuum and the pump frequency is tuned to resonantly excite
the $m=6$ biexciton state.
(b) Dashed lines are the dispersion relation of exciton-polariton
in bulk crystal.
The horizontal bars represent the exciton-photon coupled modes
with V-polarization in the film with $\thick=200\;\nm$.
The bar length is the sum of radiative and nonradiative decay rates
and the center is the resonance frequency.
(c) The H-polarized modes are plotted. Because of the breaking 
of translational invariance in the $z$ direction 
and the non-zero scattering angle,
the longitudinal excitons are also optically allowed.}
\label{fig:5}
\end{figure}
Fig.~\ref{fig:5}(a) shows the polarization-resolved
scattering spectra for thickness $\thick=200~\nm$.
The scattering angle is $\theta = 60^{\circ}$ and
the pump frequency is tuned to excite the $m = 6$ biexciton state.
Compared to the spectra for bulk crystal in Fig.~\ref{fig:4},
there are more than two peaks in the LEP-HEP frequency region.
The peak positions are different for H- and V-polarizations,
and they do not obey the angle-frequency relation for bulk crystal.\cite{inoue76}
The spectral shape can be interpreted by the exciton-photon coupled modes
\cite{ishihara04,syouji04,kojima08,bamba09radlett,bamba09crossover,ichimiya09prl}
in the thin film, which have been discussed in relation with
the radiative decay of excitons in nano-to-bulk crossover regime.
\cite{knoester92,bjork95,agranovich97,ajiki01}
Due to the breaking of translational symmetry in $z$ direction,
the lower and upper polaritons in bulk material are no longer good eigen states,
but instead we obtain the exciton-photon coupled modes with discrete energy levels
and radiative decay rates in the case of thin films.
A created biexciton spontaneously decays into these coupled modes
with emitting a photon conserving the energy and in-plane wavevector.
By using the method in Ref.~\onlinecite{bamba09crossover},
we calculated the exciton-photon coupled modes with V-polarization
in the film with $\thick = 200\;\nm$ as shown in Fig.~\ref{fig:5}(b)
and the modes with H-polarization are shown in Fig.~\ref{fig:5}(c).
The dashed lines represent the dispersion relation of exciton-polariton 
in bulk crystal, and the horizontal bars are the coupled modes
in the thin film.
The length of each bar represents the sum of radiative and nonradiative decay rates,
and the center is the resonant frequency.
Since the H-polarized modes includes the longitudinal components, 
there are also the exciton-like modes with longitudinal exciton energy.
The higher frequency parts of the scattering spectra in Fig.~\ref{fig:5}(a)
apparently reflect the structure of the coupled modes
shown in Figs.~\ref{fig:5}(b) and (c),
and the peaks in lower frequency part appear with satisfying the energy conservation.
In this way, the scattering spectra of thin films
are completely different from the bulk one,
and they depend on the film thickness, surroundings, and in-plane wavenumber
obeying the change of exciton-photon coupled modes as discussed in Ref.~\onlinecite{bamba09crossover}.
Furthermore, in contrast to the spectra for bulk crystal
in Fig.~\ref{fig:4},
the emission near the exciton resonance $\omega \simeq \wT$
can go outside the film, because of the large radiative decay rate in the thin film.
From these results, the measurement of scattering spectra of RHPS
can be considered as a powerful tool \cite{ishihara07} to observe the exciton-photon coupled modes
in nano-structured materials in addition to the previously performed
nonlinear optical responses.\cite{syouji04,ichimiya09prl}

\subsection{Entangled-photon generation\label{sec:perform}}
\begin{figure}[tbp] 
\includegraphics[width=.48\textwidth]{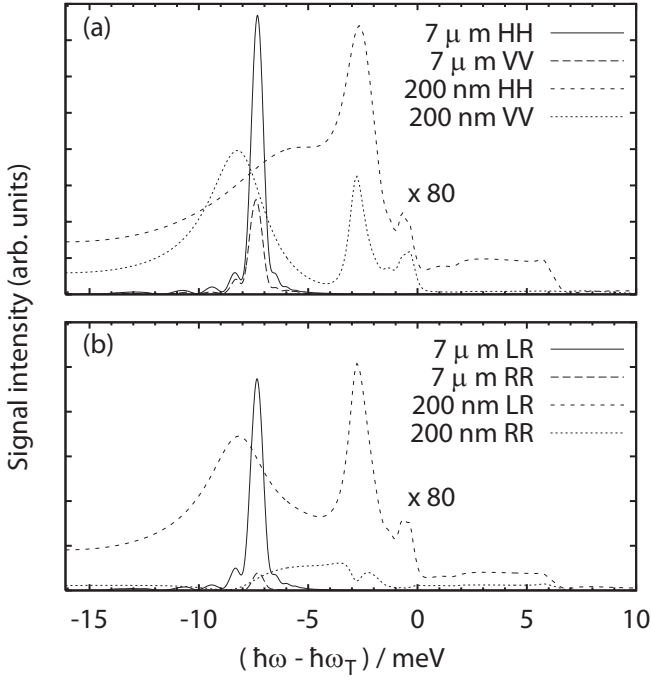}
\caption{The two-photon coincidence (signal) intensity
is plotted as a function of scattering frequency $\omega$ of one photon.
The spectra are resolved by the polarization of the two photons.
CuCl films with thicknesses $\thick=7\;\microm$ and $200\;\nm$ are considered,
and the scattering angle is $\theta=60^{\circ}$.
The same pump power is supposed for both the thicknesses.
(a) The polarization directions of two photons are resolved in H- and V-axes,
and $HH$ and $VV$ pairs are correlated.
The pairs with $HV$ and $VH$ polarizations are not generated 
from the lowest biexciton level in a film.
(b) The polarization directions are resolved for circular polarization base.
Not only the pairs of left (L) and right (R) circularly polarizations
are obtained, but $LL$ and $RR$ pairs are also generated for $\theta > 0$.
The spectra of $LR$ and $RL$ are the same, 
and those of $RR$ and $LL$ are also the same.
The spectra of 200\;nm are magnified by factor 80 in both (a) and (b).
Since the spectra are symmetric about $\win$,
only the higher frequency side $\omega > \win$
is plotted.}
\label{fig:6}
\end{figure}
Next, we discuss the entangled-photon generation by RHPS process.
Fig.~\ref{fig:6} shows polarization-resolved spectra
of two-photon coincidence measurement.
We plot only the signal intensity
$\intsignal_{\xi_1,\xi_2,\kpin=0,\kpone=(\wT/c)\sin\theta}(z_1>\thick,z_2>\thick,\win,\omega)$
as a function of scattering frequency $\omega$ of one photon
(the other photon has a frequency $2\win-\omega$).
In Fig.~\ref{fig:6}(a), the pairs with $\xi_1=\xi_2=H$ and $\xi_1=\xi_2=V$
are considered, and the pairs with different polarizations,
such as $HV$ and $VH$, have no correlation,
because the lowest biexciton level with zero angular momentum is excited.
Two film thicknesses $7\;\microm$ and $200\;\nm$ are considered,
and the parameters are the same as in Figs.~\ref{fig:4}
and \ref{fig:5}, respectively.
In the two calculation, we considered the same pump power $\Iin$,
and the ratio of signal intensities do not depend on $\Iin$.
Since the spectra are symmetrical about $\win$,
we show only the higher energy part $\omega > \win$.
As similar as the scattering spectra in Fig.~\ref{fig:4},
the intensities of $HH$ and of $VV$ are not the same in general
in the case of non-zero scattering angle.
Therefore, the ideal entanglement in Eq.~\eqref{eq:Phi+=HH+VV}
is not generally obtained, and the entangled state also 
have $RR$ and $LL$ components,
whose spectra are shown in Fig.~\ref{fig:6}(b).
Although this is a general property of bulk crystals,
the situation is different in the case of nano-to-bulk crossover regime.
As seen in Fig.~\ref{fig:6}(a), we can obtain the same signal intensities
for $HH$ and $VV$ at frequencies $\hbar\omega-\hbar\wT = -9.8$ and $-7.7\;\meV$
for $\thick = 200\;\nm$,
and the signal intensities of $RR$ and $LL$ become nearly zero at frequency
$-8.8\;\meV$ in Fig.~\ref{fig:6}(b),
while it is slightly different from the peak frequency
$-8.2\;\meV$ of $LR$ spectrum.
These results mean that the state of emitted photon pairs can be modified
by tuning the film thickness, scattering angle, and scattering frequency
in the case of thin films.
For example, at frequency $-8.8\;\meV$ for $\thick=200\;\nm$,
we can get the entangled state $(\ket{LR}+\ket{RL})/\sqrt{2}$,
while the proportions of $\ket{HH}$ and $\ket{VV}$ are not equal
as seen in Fig.~\ref{fig:6}(a),
because the polarization basis of the two photons are different for $\theta\neq0$.
On the other hand, at frequencies $-9.8$ and $-7.7\;\meV$,
we get the entangled pairs with the same $HH$ and $VV$ proportions,
while they contains $RR$ and $LL$ components.

\begin{figure}[tbp] 
\includegraphics[width=.48\textwidth]{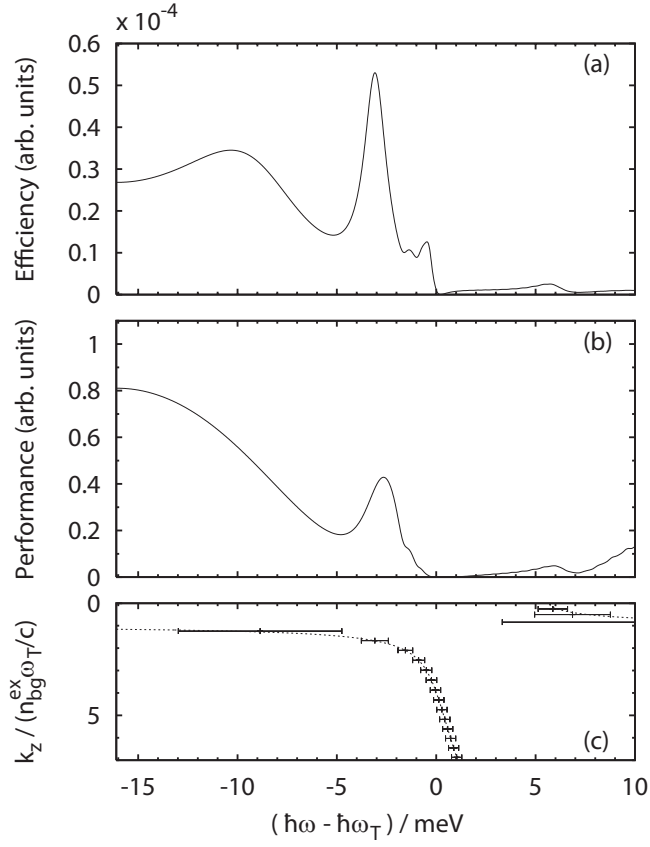}
\caption{(a) The generation efficiency (signal intensity) 
from a CuCl film with thickness $\thick=200\;\nm$ is plotted
as a function of scattering frequency $\omega$.
The biexciton state $m = 6$ is resonantly excited, 
and the scattering angle is $\theta = 0^{\circ}$.
(b) The spectrum of the corresponding performance $P$.
The generation efficiency and the performance are normalized
in the same manner as in Fig.~\ref{fig:8}.
(c) The exciton-photon coupled modes for $\theta=0^{\circ}$
in the film are plotted,
while $\theta=60^{\circ}$ in Fig.~\ref{fig:5}(b).}
\label{fig:7}
\end{figure}
Furthermore, even if the scattering angle is $\theta=0^{\circ}$, 
in contrast to the bulk case,
the scattering peaks are not at $\omega = \wT$ in general
in the case of thin films.
Therefore, the maximally entangled photon pairs are obtained
by the frequency filtering under the observation at $\theta=0^{\circ}$.
Fig.~\ref{fig:7}(a) shows the spectrum of signal intensity
(generation efficiency) obtained by a CuCl film with thickness
$\thick=200\;\nm$ for scattering angle $\theta = 0^{\circ}$.
The proportions of $HH$ and $VV$ are the same, 
and $RL$ and $LR$ paris are not emitted.
As seen in Fig.~\ref{fig:7}(a), the peaks appear not at $\omega = \win$
but close to the resonance frequencies of the exciton-photon coupled modes
shown in Fig.~\ref{fig:7}(c) (but not just at the resonance frequency
because we get weaker absorption at frequency far from $\wT$).
In this way, compared to the bulk crystals\cite{edamatsu04,oohata07}
and also to the simple quantum dots,\cite{akopian06,stevenson06,Salter2010N}
the nano-to-bulk crossover regime has a variety of degrees
of freedom to tune the generated state.

For the high-power generation of the entangled photons,
the important factors are the generation efficiency and 
also the statistical accuracy, i.e., the amount of unentangled pairs.
For a pumping beam with power $\Iin$,
the signal intensity $S \propto \intsignal$ (amount of entangled pairs) 
is proportional to ${\Iin}^{2}$, 
while the noise intensity $N \propto \intnoise$ (amount of unentangled pairs)
is proportional to ${\Iin}^{4}$,
because an unentangled pair is involved with two biexcitons.
This implies that, by increasing the pump power $\Iin$,
the $S/N$ ratio decreases in contrast to the increase in $S$.\cite{oohata07}
To evaluate the material potential
for the generation of strong and qualified entangled-photon beams,
we introduce another measure termed ``performance'' $P$
defined as the signal intensity $S$
under a certain $S/N$ ratio $\alpha$
($\Iin$ is tuned to give this ratio).
This quantity $P = S^2/\alpha N$ does not depend on $\Iin$
and reflects the material potential.
Fig.~\ref{fig:7}(b) shows the spectrum of the performance.
As comparing with Fig.~\ref{fig:7}(c), the spectrum of $P$ basically reflects the
exciton-photon couples modes.
However, since the spectrum of the noise (twice the scattering intensity)
is different from the signal one,
the Figs.~\ref{fig:7}(a) and \ref{fig:7}(b) are slightly different.
The most significant difference is the spectra around
$\omega = \win \sim \wT-16.1\meV$.
While both $S$ and $P$ mostly reflect the resonance frequency of the exciton-photon
coupled modes, the performance is maximized at $\omega = \win$,
because $P$ is strongly affected by the nonradiative damping,
which becomes smaller at that frequency.

\begin{figure}[tbp] 
\includegraphics[width=.48\textwidth]{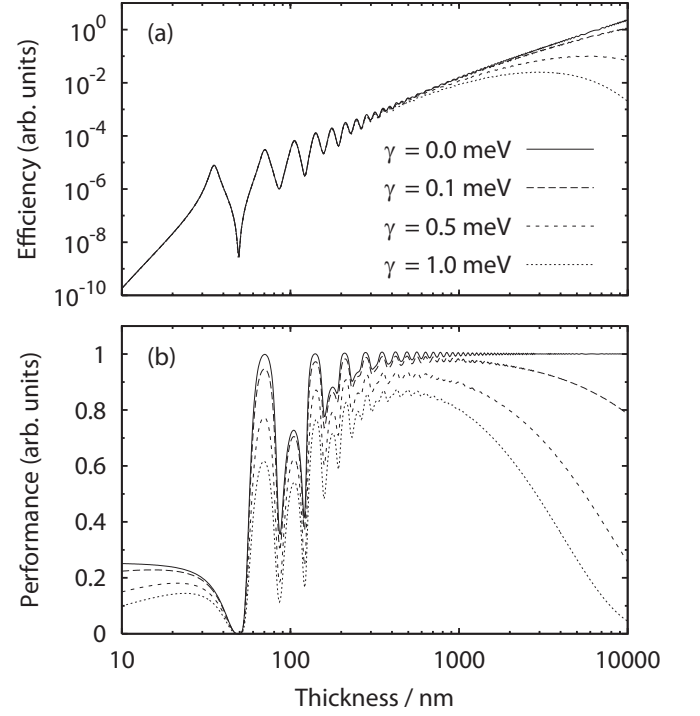}
\caption{Thickness dependences of (a) generation efficiency $S/{\Iin}^2$ 
and (b) performance $P$.
The scattering angles are $\theta=0^{\circ}$ and
the frequencies are $\omega_{1/2}=\win\pm0^+$.
To suppress the interference effect, 
outside medium is the dielectrics with $\diebgex$.
The results for $\dampex=0$, 0.1, 0.5, and $1.0\;\meV$ are plotted with
different lines.
The performance is normalized to the ideal quantity.}
\label{fig:8}
\end{figure}
Fig.~\ref{fig:8} shows the thickness dependences of (a) generation efficiency
$S/{\Iin}^2$ and (b) performance $P$.
The shapes of $P$ curves do not depend on a chosen $\alpha$,
and the maximum value is normalized to unity.
We also plot the generation efficiency with arbitrary units,
because the estimation of absolute signal intensities are sensitive to the change of
measurement conditions,
while the spectral shape and thickness dependence do not depend
on such conditions.
For simplicity, we assume that
the two scattering fields are forward and perpendicular to the film
($\theta = 0^{\circ}$) and the frequencies are $\omega_{1/2} = \win\pm 0^+$.
The pump frequency is tuned to the two-photon absorption in bulk material.
The results for nonradiative decay rates
$\dampex = 0$, 0.1, 0.5, and $1.0\;\mathrm{meV}$
are plotted with different lines.
In order to suppress the oscillation due to the interference as seen in
Fig.~\ref{fig:3}, we suppose that the CuCl film exists in a dielectric medium
with $\diebgex$.
The results for the film in vacuum will be shown in Fig.~\ref{fig:9}.
In Fig.~\ref{fig:8}, the oscillating behavior in the nanometer thickness range 
is due to the biexciton confinement and the modification of the energy structure
of exciton-photon coupled modes.
The RHPS effectively occurs when
the resonance energy of the coupled mode is
just equal to half the biexciton energy.
The maximum performance shown in Fig.~\ref{fig:8}(b) is the ideal quantity,
and it only depends on measurement conditions, 
such as resolutions of angle and frequency, but not on material parameters.

As seen in Fig.~\ref{fig:8}(a) and also in 
Ref.~\onlinecite{savasta99ssc}, the optimal thickness for
generation efficiency depends on $\dampex$, 
and it is on the order of micrometers or more for CuCl crystals.
However, as seen in Fig.~\ref{fig:8}(b),
the performance significantly decreases from the ideal value
at a thickness of micrometers for nonzero $\dampex$,
because the nonradiative decay easily increases the amount of 
unentangled pairs.
Therefore, when we use bulk crystals, the generation efficiency
(generation probability for one pump pulse)
is limited by a desired statistical accuracy ($S/N$ ratio).\cite{oohata07}
However, at a thickness from 50 to 1000\;nm,
as expected, 
a nearly ideal performance can be obtained at particular thicknesses
even if $\dampex$ is nonzero.
This is because the radiative decay is dominant
owing to the exciton superradiance,\cite{bamba09crossover,ichimiya09prl}
and the entangled pairs can go outside the film
without decreasing the amplitude.
Therefore, thin films generally show a high performance,
and this rapid decay is also desired
for the high-repetition excitation,
which also recovers the signal intensity while maintaining the $S/N$ ratio.
\cite{oohata07}

\subsection{With DBR cavity\label{sec:PRL}}
\begin{figure}[tb] 
\includegraphics[width=.48\textwidth]{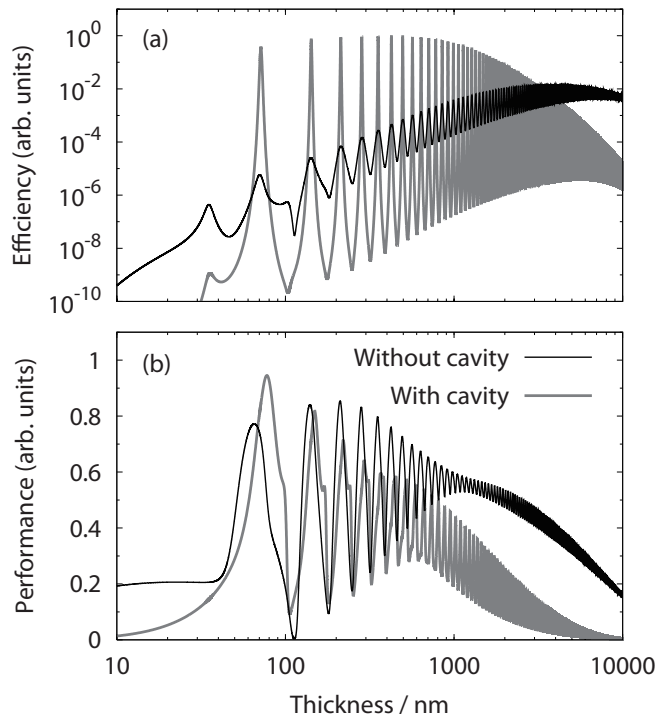}
\caption{Thickness dependences of (a) generation efficiency $S/{\Iin}^2$ 
and (b) performance $P$.
The black lines represent the results for a bare CuCl film existing in vacuum.
The scattering frequencies are the same as that of pumping beam,
and scattering is forward with $\theta=0^{\circ}$.
The gray lines represent the results for DBR cavity embedding a CuCl layer
shown in Fig.~\ref{fig:2}(b).
The mode frequency of the optical cavity is tuned to $\wT$,
and the scattering is backward with $\theta=180^{\circ}$.
In both cases, $\dampex = 0.5\;\meV$.
The performance is normalized in the same manner as in Fig.~\ref{fig:8}.}
\label{fig:9}
\end{figure}

Although a good performance is obtained at a thickness of
hundreds nanometers, the generation efficiency of such thin films is much lower 
than that of bulk crystals as seen in Fig.~\ref{fig:8}(a),
and a strong pump power is required to achieve a sufficient signal intensity
at such thickness range.
While the superradiant excitons maintain the large nonlinearity (excitonic component),\cite{ichimiya09prl}
this low efficiency simply comes from the small thickness (interaction volume).
This problem can be overcome
by using an optical cavity
in the strong-coupling regime, 
because we can control both the interaction volume and radiative decay rate
using two parameters: quality factor (Q-factor) of cavity and thickness of CuCl.
This aspect is different from the simple semiconductor microcavity,
in which the interaction volume and radiative decay rate are
respectively enhanced in strong- and weak-coupling regimes.

Although a high generation efficiency can be achieved by using a high-Q cavity,
we have to simultaneously realize a rapid radiative decay of entangled photons
inside the cavity. Therefore, we consider a low-Q cavity as reported in Ref.~\onlinecite{oohata08}, namely, a CuCl layer in a DBR cavity composed 
of $\mathrm{PbF_2}$ and $\mathrm{PbBr_2}$
as seen in Fig.~\ref{fig:2}(b).
Here, since the translational invariance is broken at the thickness of nanometers,
the generated photons can go forward and also backward in contrast to the bulk case.
Therefore, we suppose a high reflectance on the transmission side
to suppress the leakage of entangled photons, and we focus on the backward emission.
The DBR cavity is considered by the background dielectric function
$\diebg(\vr,\omega)$ in Eq.~\eqref{eq:Maxwell-E-Jz-Pex}.
The refractive indexes of $\mathrm{PbF_2}$ and $\mathrm{PbBr_2}$
are 1.86 and 2.95, respectively.
The gray lines in Fig.~\ref{fig:9} represent the backward emission
from the cavity structure,
where the cavity mode frequency is tuned to $\wT$,
$\dampex = 0.5\;\mathrm{meV}$,
and the periods of the incident and transmitted sides are 4 and 16,
respectively (Q-factor is 50).
This system corresponds to the weak bipolariton regime\cite{ajiki07,oka08}
(but the strong-coupling regime of excitons and photons),
where the energy splitting between polariton and biexciton levels is small
compared with their broadening. 
This situation is in contrast to that in
Refs.~\onlinecite{ajiki07} and \onlinecite{oka08},
where the strong enhancement of 
entangled-photon generation from a quantum well in a high-Q cavity 
has been discussed on the basis of the biexcitonic cavity-QED picture or the strong bipolariton picture.
As shown in Fig.~\ref{fig:9}(a),
since biexcitons are effectively created,
the generation efficiency is significantly enhanced at a thickness of nanometers,
and it is larger than the maximum value
in the case of bare CuCl film existing in vacuum (black line).
The enhancement also occurs when the polariton energy
(exciton-photon coupled mode) is equal to half the biexciton energy,
which is consistent with the results in Refs.~\onlinecite{ajiki07}
and \onlinecite{oka08}. 
Compared with Fig~\ref{fig:8}(a),
the period of the oscillation is doubled,
because the RHPS involving biexcitons 
with odd-parity center-of-mass motion
is forbidden in the one-sided optical cavity.
On the other hand, as shown in Fig.~\ref{fig:9}(b),
at a thickness of micrometers,
the performance is suppressed compared with that of the bare CuCl film.
This is because of the multiple reflections inside the cavity,
and the scattered fields non-radiatively decay during the propagation.
In contrast, at the thickness of nanometers, particularly at 80\;nm,
the performance almost maintains the ideal quantity.
This is due to the enhancement of the radiative decay rate by exciton superradiance,
and the enhancement of generation efficiency is simultaneously obtained
by the cavity effect in the strong-coupling regime.

\begin{figure}[tb] 
\includegraphics[width=.48\textwidth]{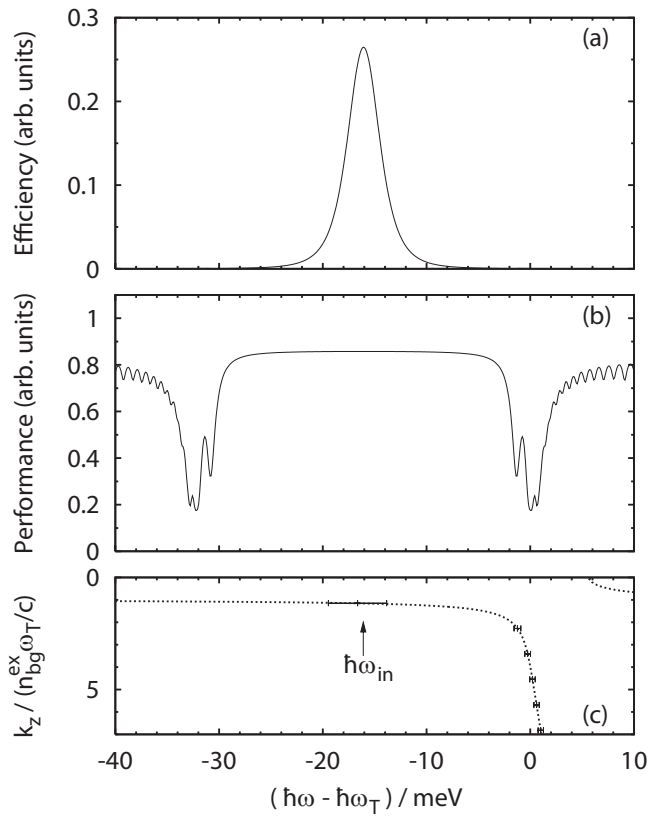}
\caption{(a) The generation efficiency (signal intensity) 
is plotted as a function of scattering frequency $\omega$.
A CuCl film with thickness $\thick=72\;\nm$ is embedded
in the DBR cavity considered in Fig.~\ref{fig:9},
and the other parameters are also the same.
(b) The spectrum of the corresponding performance $P$.
(c) The exciton-photon coupled modes in the film are plotted
in the same manner as in Fig.~\ref{fig:5}(b).}
\label{fig:10}
\end{figure}
Finally, in Figs.~\ref{fig:10}(a) and (b),
we show the spectra of generation efficiency and performance, respectively,
in the case of CuCl film with thickness $d = 72\;\nm$ embedded in
the DBR cavity discussed above.
This thickness is chosen to achieve a high generation efficiency,
while the performance at $\omega = \win$ is 0.86,
which is smaller than the maximum value in Fig.~\ref{fig:9}(b).
However, compared to the thin film without the cavity,
the generation efficiency is significantly increased
and the high performance is successfully maintained
due to the rapid radiative decay.
Further, while the pump frequency $\win$ is assumed to
the two-photon absorption frequency in bulk CuCl in Figs.~\ref{fig:10}(a) and (b),
we have numerically checked that,
when $\win$ is correctly tuned to the eigen frequency
of a confined biexciton mode,
the generation efficiency is enhanced more dramatically
with maintaining the high performance in the case with DBR cavity.

In Fig.~\ref{fig:10}(c), the exciton-photon coupled modes
are plotted with horizontal bars, 
and we can find that one mode with high radiative decay rate
exists close to the pump frequency $\win$
(two-photon absorption frequency of biexcitons).
This mode corresponds to the lower cavity polariton,
and the strong electric field in the cavity
enhances the generation efficiency of biexcitons
due to the cavity-induced double resonance.
\cite{Ishihara1997APL}
Furthermore, the generated entangled excitons rapidly decay into photons
through this polariton mode,
which ensures the high performance at the same time.

In this way, by using an optical cavity embedding a CuCl layer
with a thickness of nanometers,
we can obtain high efficiency and performance simultaneously.
To avoid the leakage of generated photons, the reflectance on the transmission side
should be high enough, but that on the incident side should not be high
to obtain a rapid radiative decay rate.
Once we choose a cavity structure,
we can numerically determine the optimal thickness of CuCl layer,
in which an exciton-photon coupled mode has half the biexciton frequency,
a rapid radiative decay rate, and also large exciton component
(large nonlinearity) to achieve high performance and generation efficiency.
Such a coupled mode is an unique feature in the nano-to-bulk crossover regime.

\section{Summary} \label{sec:summary}
We have developed a theoretical framework for the investigation
of the biexciton-involved
RHPS based on the QED theory for excitons.\cite{bamba08qed}
Compared to the previous theories,\cite{savasta99prb,savasta99ssc}
our method can be applied to nano-to-bulk crossover regime
because we explicitly consider the center-of-mass motion of exciton.
Further, we can discuss the polarization correlation of entangled pairs
and the surroundings of the excitonic layer, such as the DBR cavity structure.
While we considered CuCl films in actual calculation,
by treating several relative exciton levels
and by correctly calculating the center-of-mass wavefunctions
of excitons confined in finite crystal
including the effect of dead layer,\cite{agranovich84}
our theoretical framework can be applied to other materials in principle.
Further, by correctly treat the modification of relative motion of excitons
and biexcitons strongly confined in nano crystals
and also the Pauli's exclusion principle,
our framework can be extended for the investigation of single quantum dot,
the deterministic generation of entangled photons.

We have calculated the scattering spectra of RHPS
from CuCl films with bulk-like and submicron thicknesses.
For the bulk-like thickness, the four peaks called $\mathrm{M_T}$, 
$\mathrm{M_L}$, LEP and HEP are reproduced.
On the other hand, the scattering spectra of the thin film
are significantly modified from the bulk ones,
and we found that they reflect the exciton-photon coupled modes
in the thin film.
\cite{syouji04,kojima08,ichimiya09prl,bamba09radlett,bamba09crossover}
Therefore, the RHPS measurement is also useful to
observe the exciton-photon coupled modes in nano-structured materials \cite{ishihara07}
as well as the four-wave mixing\cite{ishihara02jul,kojima08,ichimiya09prl}
and the two-photon excitation scattering.\cite{syouji04}
We also found that semiconductor thin films have much degrees of freedom 
to control generated states of entangled photon pairs.

In addition to the signal intensity of entangled-photon generation,
we also discuss the performance of the material structure
by considering the noise intensity from independent two biexcitons.
Although the thickness dependence of signal intensity
has a maximum value at particular thickness
determined by nonradiative decay rate of excitons,
\cite{savasta99prb,savasta99ssc}
a high performance is generally obtained at thickness
of nanometers due to the rapid radiative decay of excitons.
However, the generation efficiency of such thin films
is much weaker compared to the bulk one.
We have demonstrated that, by using a DBR cavity embedding an excitonic layer
in the nano-to-bulk crossover regime,
the generation efficiency can be enhanced
while maintaining the high performance.

For the pursuit of high-power and high-quality but probabilistic
generation of entangled photons, which is essential for
the next-generation technologies of fabrication
and chemical reaction,\cite{Kawabe2007OE}
the biexciton-involved RHPS shows a quite high generation efficiency
compared to that of PDC method.
From the viewpoint of the quality of generated entangled pairs,
the RHPS method can show a high performance
and a high generation efficiency simultaneously
by using and an optical cavity embedding a CuCl nano-layer.
Further, it has much degrees of freedom to control the generated states
of entangled photons.
We believe that our results make a breakthrough
in high-power and high-quality entangled-photon generation.

\begin{acknowledgments}
The authors are grateful to K.~Edamatsu, G.~Oohata, H.~Ajiki, and H.~Oka for helpful discussions.
This work was partially supported by the Japan Society for the Promotion
of Science (JSPS): 
a Grant-in-Aid for Creative Science Research 17GS1204 (2005)
and Grant No.~19-997.
\end{acknowledgments}


\end{document}